\documentclass[fleqn,usenatbib,useAMS]{mnras}
\usepackage{xcolor}

\usepackage{lipsum}
\usepackage{comment}
\usepackage{graphicx}
\usepackage{amsmath}
\usepackage{amssymb}
\usepackage[super]{nth}
\usepackage[T1]{fontenc}
\usepackage{ae,aecompl}

\usepackage{newtxtext,newtxmath}

\DeclareRobustCommand{\VAN}[3]{#2}
\let\VANthebibliography\thebibliography
\def\thebibliography{\DeclareRobustCommand{\VAN}[3]{##3}\VANthebibliography}

\title[SILCC VII]{SILCC VII - Gas kinematics and multiphase outflows of the simulated ISM at high gas surface densities}

\author[T.-E. Rathjen et al.]{
	Tim-Eric Rathjen,$^{1, 2}$\thanks{E-mail: rathjen@ph1.uni-koeln.de}
	Thorsten Naab,$^{1}$
	Stefanie Walch,$^{2}$
	\newauthor	
	Daniel Seifried,$^{2}$
	Philipp Girichidis$^{3}$
	\& Richard W\"unsch$^{4}$
	\\
	$^{1}$Max Planck Institute for Astrophysics, Karl-Schwarzschild-Str. 1, 85748 Garching, Germany\\
	$^{2}$I. Physikalisches Institut, Universit\"at zu K\"oln, Z\"ulpicher Str. 77, 50937 K\"oln, Germany\\
	$^{3}$Universit\"at Heidelberg, Zentrum f\"ur Astronomie, Institut f\"ur Theoretische Astrophysik, Albert-Ueberle-Str. 2, 69120 Heidelberg, Germany\\
	$^{4}$Astronomical Institute of the Czech Academy of Sciences, Bo\v{c}n\'{i} II 1401, 141 00 Prague, Czech Republic
}

\date{Accepted XXX. Received YYY; in original form ZZZ}

\pubyear{2023}

\begin{document}
\label{firstpage}
\pagerange{\pageref{firstpage}--\pageref{lastpage}}
\maketitle

\begin{abstract}
We present magnetohydrodynamic (MHD) simulations of the star-forming multiphase interstellar medium (ISM) in stratified galactic patches with gas surface densities $\Sigma_\mathrm{gas} =$ 10, 30, 50, and 100 $\mathrm{M_\odot\,pc^{-2}}$. The \textsc{silcc} project simulation framework accounts for non-equilibrium thermal and chemical processes in the warm and cold ISM. The sink-based star formation and feedback model includes stellar winds, hydrogen-ionising UV radiation, core-collapse supernovae, and cosmic ray (CR) injection and diffusion. The simulations follow the observed relation between $\Sigma_\mathrm{gas}$ and the star formation rate surface density $\Sigma_\mathrm{SFR}$. CRs qualitatively change the outflow phase structure. Without CRs, the outflows transition from a two-phase (warm and hot at 1 kpc) to a single-phase (hot at 2 kpc) structure. With CRs, the outflow always has three phases (cold, warm, and hot), dominated in mass by the warm phase. The impact of CRs on mass loading decreases for higher $\Sigma_\mathrm{gas}$ and the mass loading factors of the CR-supported outflows are of order unity independent of $\Sigma_\mathrm{SFR}$. Similar to observations, vertical velocity dispersions of the warm ionised medium (WIM) and the cold neutral medium (CNM) correlate with the star formation rate as $\sigma_\mathrm{z} \propto \Sigma_\mathrm{SFR}^a$, with $a \sim 0.20$. In the absence of stellar feedback, we find no correlation. The velocity dispersion of the WIM is a factor $\sim 2.2$ higher than that of the CNM, in agreement with local observations. For $\Sigma_\mathrm{SFR} \gtrsim 1.5 \times 10^{-2}\,\mathrm{M}_\odot\,\mathrm{yr}^{-1}\,\mathrm{kpc}^{-2}$ the WIM motions become supersonic.
\end{abstract}


\section{Introduction}\label{sec:intro}
Galactic outflows are unambiguously important in shaping the evolution of galaxies and could be a major agent in the regulation of star formation \citep{Veilleux2005}. Understanding the multiphase nature, the kinematic structure, and the magnitude of galactic outflows is imperative to inform subgrid models for larger-scale cosmological simulations and successfully interpret observations. 
One driver of galaxy-scale outflows in active galaxies are active galactic nuclei (AGN) \citep[see e.g.][and references therein]{Harrison2018}. For star-forming galaxies, the origin of galactic outflows is the multiphase interstellar medium (ISM) \citep[see for a review][]{Veilleux2005}. It is established from observations and numerical simulations that galactic outflows are multiphase, as is the multiphase ISM \citep[see][for a review on the phase structure of the ISM]{Ferriere2001} out of which they are launched into the circumgalactic medium (CGM). Most of the mass in galactic outflows of star-forming galaxies is carried in the warm gas phase ($T \sim 10^4\,\mathrm{K}$) while most of the energy and metals leave the ISM through channels of hot gas ($T \geq 10^6\,\mathrm{K}$) \citep{Veilleux2005, Kim2020a}. The creation of a volume-filling hot gas phase through overlapping supernova (SN) explosions is identified as the source of fast-moving, ballistic, and warm-hot galactic winds \citep{Li2020}. However, SN-driven outflow models seem to fail to create cold gas ($T < 300\,\mathrm{K}$, CNM) or even molecular outflows with mass loading factors high enough, as they are observed for local star-forming galaxies, as well as for active galaxies \citep{Fluetsch2019, Lutz2020, Veilleux2020, DiTeodoro2019}.
Cosmic rays (CRs) might be the missing feedback channel through which a colder, smoother, and slower-moving outflow can be lifted away from the midplane ISM \citep{Girichidis2018, Rathjen2021}. CRs can be accelerated in the shock fronts of SN remnants by diffusive shock acceleration \citep{Bell1978} with energies approximately $10\,\mathrm{per\,cent}$ of the SN energy \citep[$\sim 10^{50}\,\mathrm{erg}$,][]{Helder2012, Ackermann2013}. The energy density in the ISM of those relativistic charged particles at mostly $\sim \mathrm{GeV}$ energies \citep{Strong2007, Nava2013} is. comparable to the thermal, magnetic and kinetic energy in the local ISM \citep{Draine2011}. Once accelerated, CRs are transported via streaming, adiabatic advection, and anisotropic diffusion along magnetic field lines through the ISM. Energy losses in low-density environments are small \citep{Pfrommer2017}, which allows CRs to establish a long-lasting large-scale vertical pressure gradient in the ISM through diffusion. This additional nonthermal CR pressure gradient might be the channel through which cold galactic outflows can be driven. However, the number of numerical simulations of the star-forming ISM that consider CR transport in their models is still limited \citep[see e.g.][]{Simpson2016, Pakmor2016, Pakmor2017, Girichidis2018, Rathjen2021, Hopkins2021, Hopkins2023, Simpson2023, Armillotta2022}.

Global galaxy simulations indicate increasing dispersions of cold and warm gas velocities in regions of higher star formation rates \citep{Ejdetjarn2022, Jimenez2022}. A trend that is also predicted by analytical work \citep{Krumholz2016, Krumholz2018}. The exact reason for the increase in ISM turbulence with increased star-formation activity has not been unambiguously determined. Two possible candidates are currently being discussed in the community. One is stellar feedback, which can drive galactic winds through overlapping superbubbles from SN explosions, and create outward expanding HII and hot stellar wind bubbles \citep[see][for a review]{Naab2017}. The other candidates for ISM turbulence are gravitational instabilities and collapse, fed by radial transport of gas through the galactic disk, as well as gas accretion out of the CGM onto the galactic disk. Observations of galactic outflows in the warm ionised gas phase (WIM) \citep[e.g.][]{Genzel2011, Zhou2017, Ubler2019} tend to favour the unified model of ISM turbulence driven by galactic instabilities \citep{Krumholz2018} over a starburst-driven model for turbulence in the ISM and galactic outflows \citep[see e.g.][]{Xu2022, Avery2021, Chu2022a}. However, the choice of the observational tracer can have a strong impact on the obtained results. The velocity dispersion of galaxies at peak star formation may be vastly overestimated when using ionised gas as a proxy for the total gas mass \citep{Girard2021}. This emphasises the need for numerical models of resolved multiphase ISM and galactic outflows to check against observational estimates. 

Understanding the evolution of the ISM and the importance of galactic outflows has been the focal point of many recent numerical studies of varying scales. Stratified box simulations (like the ones presented in this work) have proven to be a potent tool to unravel the plethora of ISM processes and their highly nonlinear interactions. The upside of being able to go to high enough spatial and temporal resolution to resolve the different processes comes with the cost of being constrained to a galactic patch and not being able to model a full galactic context.
The effects and relative importance of the various stellar feedback processes have been subsequently studied in previous iterations of the \textsc{silcc} project, which are executed with the magnetohydrodynamic (MHD) adaptive mesh refinement (AMR) code \textsc{flash} \citep{Fryxell2000}. The way in which SNe shape the chemical evolution of the ISM and drive galactic outflows is studied in \citet{Walch2015a} and \citet{Girichidis2016a}. More complexity to the models is added in follow-up studies of the self-regulation of massive star formation through the inclusion of stellar winds \citep{Gatto2017} and hydrogen-ionising radiation \citep{Peters2017}. A detailed analysis of the effects of a magnetised ISM is presented in \citet{Girichidis2018a}. A first look at the impact of CRs in various combinations of the aforementioned stellar feedback processes in solar neighbourhood conditions is shown in \citet{Rathjen2021}. The \textsc{silcc} project simulations in solar neighbourhood conditions have emphasised the need to include early (i.e. ionising radiation and stellar winds) and late (i.e. SNe) stellar feedback processes in unison in order to model a realistic and self-consistent ISM.
The \textsc{tigress} simulation suite \citep{Kim2017, Kim2018} uses the MHD adaptive mesh refinement AMR code \textsc{athena} to model an SN-regulated ISM with a sink particle approach, grain photoelectric heating of far-ultraviolet radiation (without radiative transfer), runaway stars, and a model for the effects of galactic shear. Heating/cooling and photochemistry \citep{Kim2023} and adaptive ray tracing for ionising UV radiation \citep{Kim2023a} have recently been added to the \textsc{tigress} model. Stellar winds and the transport of CRs are not considered. They achieve two-phase galactic outflows (warm: $T \sim 10^4\,\mathrm{K}$ and hot: $T \gtrsim 10^6\,\mathrm{K}$), which regulate star formation with mass loading factors following observational estimates \citep{Kim2020a, Kim2020b}. The multiphase ISM study by \citet{Butler2017} uses the AMR MHD code \textsc{ramses}. It includes the important early stellar feedback channel of hydrogen-ionising extreme ultraviolet radiation (EUV) of massive stars (but without stellar winds) in addition to SNe and chemistry and emphasises the importance of EUV radiation in reducing star formation efficiency, similar to earlier results from the \textsc{silcc} project \citep{Peters2017, Rathjen2021}. \citet{Kannan2020} use the moving mesh code \textsc{arepo-rt} to model the ISM in low gas surface densities ($10\,\mathrm{M_\odot}\,\mathrm{pc}^{-2}$) while accounting not only for ionising radiation from massive stars, but also for the effects of radiation pressure, and conclude that photoionisation and the creation of HII regions have the strongest effect on reducing the star formation rate by more than a factor of $\sim 2$ compared to simulations that omit EUV radiation. Simulations with the AMR MHD code \textsc{enzo} by \citet{Li2017} focus on an ISM regulated by SN and galactic outflows for systems with gas surface densities up to $150\,\mathrm{M_\odot\,\mathrm{pc}^{-2}}$ and report a decreasing total mass loading factor with increasing gas surface density \citep[see also][]{Li2020}.
Moving to scales of isolated galaxies or cosmological zoom-in simulations (usually at the cost of lesser temporal/spatial resolution and/or physical complexity or only for systems very low in total mass) enables, among others, the study of the galactic context on the ISM which is especially important to accurately capture the replenishment of the star-forming gas reservoir through fountain flows, the driving of ISM turbulence triggered from balancing gravitational energy via radial mass transport through the galactic disk, and the influence of the CGM \citep[see e.g.][]{Muratov2015, Hu2017, Emerick2018, Agertz2020, Gutcke2021, Smith2021, Pandya2021}.

With this work, we want to study the multiphase galactic outflows and related gas kinematics that originate from the star-forming ISM over a range of initial gas surface densities, $\Sigma_\mathrm{gas} = 10 - 100 \,\mathrm{M_\odot\,pc^{-2}}$. Our goal is to quantify the phase structure of the outflows and to narrow down the possible source of turbulence in the ISM. To do so, we run state-of-the-art MHD simulations of stratified galactic patches and include all major stellar feedback processes, most notably on-the-spot radiative transfer for hydrogen-ionising radiation from massive stars, momentum injection from stellar winds, and the acceleration and transport of CRs in the advection-diffusion limit. This paper is structured as follows. In Section \ref{sec:numerics}, we briefly introduce the \textsc{silcc} simulation framework and the numerical realisations of the most important ISM processes and explain the initial conditions of our simulations; in Section \ref{sec:TheSimulatedInterstellarMedium}, we present a general overview of the simulations and their qualitative properties; in Section \ref{sec:StarFormationAndOutflows}, we quantify the star formation and galactic outflow properties and put an extended focus on the phase structure of the outflow (Section \ref{sec:PhaseStructureOfTheOutflow}); we analyse the gas kinematics within the WIM and CNM in Section \ref{sec:kinematics} and close the results section in Section \ref{sec:ISMstructure} with a short analysis of the multiphase ISM in our models. We continue with a discussion of our results regarding the characteristics of galactic outflow, the role of CRs, and the origin of the velocity dispersion in Section \ref{sec:discussion}. We also discuss caveats and possible improvements to our models. We conclude our analysis in Section \ref{sec:conclusion}. 
In the Appendix, we compare two different methods of determining the velocity dispersions in our simulations, present further analysis on a test model with stellar feedback being turned off, and give the numerical results plotted in our figures in tabulated form.

\section{Numerical methods and simulation setup}\label{sec:numerics}

The stratified disk ISM simulations of a galactic patch in an elongated box presented in this paper are part of the \textsc{silcc} simulation framework \citep{Walch2015a, Girichidis2016a, Gatto2017, Peters2017, Girichidis2018a, Rathjen2021} and closely follow the setup and numerical methods described in \citet{Rathjen2021}. We use an updated version of the AMR MHD code \textsc{flash} v4.6 \citep{Fryxell2000, Dubey2009}. We briefly give an overview of the setup and realisation of certain physical processes for the simulations presented here and refer the reader to \citet{Rathjen2021} and the references therein for more details.

\textbf{Star formation} is realised with a Lagrangian subgrid sink particle approach \citep{Federrath2010, Gatto2017}. Gas above a threshold density $\rho_\mathrm{thr} = 2.1 \times10^{-21}\,\mathrm{g\,cm^{-3}}$ and satisfying additional criteria will form a sink particle. These criteria demand the gas to be Jeans-unstable, to exist in a gravitational potential minimum, and to be in a converging flow. After a sink particle is formed it can continue to accrete more gas with an accretion radius of $r_\mathrm{accr} = 3 \times dx \approx 11.7\,\mathrm{pc}$, if all of the above conditions are met\footnote{Please note that when a cell fulfils the star formation criteria and is above $\rho_\mathrm{thr}$, not all gas of the cell will be accreted but only the amount of gas above the threshold density.}. For every $120\,\mathrm{M_\odot}$ of accreted gas, we form one massive star, which we sample from a Salpeter IMF in the mass range of $8 - 120\,\mathrm{M_\odot}$ \citep{Gatto2017}. The leftover mass is assumed to go into low-mass stars, which we do not track explicitly. The evolution of individual massive stars is individually followed by Geneva stellar evolution tracks \citep{Ekstrom2012}. The total amount of feedback of a sink particle is integrated over all massive stars within the sink particle and then injected into the surrounding medium. The N-body dynamics of the sink particles is integrated with a Hermite integrator of \nth{4} order \citep{Dinnbier2020}.

At the end of the lifetime of each massive star, we realise core-collapse \textbf{Supernovae} explosions by injecting $10^{51}\,\mathrm{erg}$ thermal energy into a region with $r_\mathrm{inj} = 3 \times dx \approx 11.7\,\mathrm{pc}$. If local gas densities within the injection region exceed the Jeans mass and we cannot resolve the Sedov-Taylor phase with at least three grid cells, we switch to momentum injection to prevent overcooling \citep{Gatto2015}. In the case of momentum injection, the chemical composition of the gas remains unchanged and the thermal energy in the injection region is set to the mean thermal energy value within it. We do not include metal enrichment.

\textbf{Stellar winds} are a source of early stellar feedback and are realised in our model through momentum injection into a region of the same volume as the SN injection region ($r_\mathrm{inj} = 3 \times dx \approx 11.7\,\mathrm{pc}$) \citep{Gatto2015}. Terminal wind velocities and stellar wind mass loss rates are interpolated from Geneva tracks starting with the zero-age main sequence stage throughout the Wolf-Rayet phase \citep{Ekstrom2012}. Metal enrichment from stellar winds is not included.

\textbf{Ionising radiation} from massive stars is responsible for creating HII regions. Furthermore, the chemical state of the ISM is strongly affected and governed by radiation. In our models, we incorporate the radiative transfer tool \textsc{TreeRay} \citep{Wunsch2018, Wunsch2021} in a configuration using the on-the-spot approximation \citep{Osterbrock1988}. We include one radiation energy band of extreme ultraviolet photons (EUV) with $E_\gamma > 13.6\,\mathrm{eV}$. From the stellar evolution tracks \citep{Ekstrom2012}, we also obtain the bolometric luminosity of each given massive star, as well as its effective temperature. This information, however, does not directly give the EUV luminosity and the amount of EUV photons emitted by that star. We, therefore, approximate the fraction of EUV photons of a massive star by computing the fraction of EUV photons in a black body spectrum with the same effective temperature as the star in question. This approximation slightly underestimates the amount of EUV photons by a few per cent\footnote{Using a single stellar population synthesis model \citep{Bruzual2003}, the ratio of EUV luminosity to bolometric luminosity for a star with an effective temperature of $T_\mathrm{eff} \approx 5 \times 10^4\,\mathrm{K}$ yields $\frac{L_\mathrm{EUV}}{L_\mathrm{bol}} \approx 0.849$, whereas for a black body spectrum, the ratio would be $\frac{L_\mathrm{EUV}}{L_\mathrm{bol}} \approx 0.769$.}. We choose this approach instead of modelling stars directly with a stellar atmosphere from single stellar population synthesis models in order to keep consistency within our stellar evolution model which accounts for the stellar mass loss rates and the total bolometric luminosities.
We accumulate the EUV photons of all massive stars of a sink particle and inject the ionising photons into the cell where the sink particle sits and propagate the radiation along 48 rays normal to an equal area isolatitude pixelation of a sphere calculated with the \textsc{HEALPix} algorithm \citep{Gorski2005}.
An important advantage of \textsc{TreeRay} is its use of the oct-tree structure and the backward radiative transfer approach. Instead of propagating the radiation from each source (i.e. the sink particles) towards each computational cell, the propagation is traced backwards from each cell towards the sources. This ensures that the computational cost does not increase with the number of radiation sources.

\textbf{Self-gravity} is included with an oct-tree-based solver for Poisson equations \citep{Barnes1986, Wunsch2018}. Additionally, we add an external potential to model the gravitational influence of the old stellar population with a stellar surface density of $\Sigma_\star = 30\,\mathrm{M_\odot\,pc^{-2}}$ and vertical scale height $H_\star = 300\,\mathrm{pc}$, as well as a potential from an NFW dark-matter profile \citep{Navarro1996} with $R_\mathrm{vir} = 200\,\mathrm{kpc}$, concentration parameter $c=12$ and distance $R = 8\,\mathrm{kpc}$ from the galactic centre. We do not scale the magnitude of the external gravitational potential with $\Sigma_\mathrm{gas}$.

\textbf{Chemistry, heating and cooling} processes are taken into account with a time-dependent non-equilibrium chemical network that includes radiative heating and cooling \citep{Nelson1997, Glover2007, Walch2015a}. We explicitly track the evolution of H, H$+$, H$_2$, C$^+$, CO, and e$^-$. For gas above $T > 10^4\,\mathrm{K}$, we assume collisional ionisation equilibrium and apply tabulated cooling rates from \citet{Gnat2012}. Heating processes include photoelectric heating of dust and polycyclic aromatic hydrocarbons (PAH) and cosmic ray ionisation, as well as changes in the thermal energy of various chemical processes, such as collisional- and photo-dissociation of H$_2$, among others. We include a background far-ultraviolet (FUV) interstellar radiation field (ISRF) to account for photoelectric heating. The strength of the ISRF scales with $\Sigma_\mathrm{gas}$ from $G_0 = 1.7\,\mathrm{to\,}42.7$ \citep{Draine1978}. Furthermore, the ISRF is attenuated by dust and the self-shielding of H$_2$ and CO. The optical depths and the respective column densities are calculated using \textsc{TreeCol} \citep{Clark2012, Wunsch2018}. This leads to an effective $G_\mathrm{eff} = G_0 \times \exp (-2.5 A_\mathrm{V})$, with visual extinction $A_\mathrm{V}$. The photoelectric heating rate is then given by (\citet{Bakes1994, Bergin2004}): \begin{align}
    \Gamma_\mathrm{pe} = 1.3 \times 10^{-24}\,\epsilon\,G_\mathrm{eff}\,n\,\mathrm{erg\,s^{-1}\,cm^{-3}},
\end{align}with the photoelectric heating efficiency, $\epsilon$, given by (\citet{Bakes1994, Wolfire2003})

All models assume solar metallicity with abundances taken from \citet{Sembach2000}. We impose a constant dust-to-gas ratio of 1 per cent.

We include \textbf{Magnetic fields}, as well as injection and propagation of \textbf{Cosmic rays} by advection and diffusion. CRs are included as a separate relativistic fluid and are dynamically coupled to the MHD equations as an additional pressure source term. The modified MHD equations read:
\begin{align}
\frac{\partial\rho}{\partial t} &+ \nabla \cdot (\rho\mathbf{v}) = 0\\
\frac{\partial\rho\mathbf{v}}{\partial t} &+ \nabla \cdot \left(\rho \mathbf{v}\mathbf{v}^\mathrm{T} - \frac{\mathbf{B}\mathbf{B}^\mathrm{T}}{4\pi}\right) + \nabla P_\mathrm{tot} = \rho\mathbf{g} + \dot{\mathbf{q}}_\mathrm{sn}\\
\frac{\partial e}{\partial t} &+ \nabla \cdot \left[\left(e + P_\mathrm{tot}\right)\mathbf{v} - \frac{\mathbf{B}\left(\mathbf{B} \cdot \mathbf{v}\right)}{4\pi}\right] \notag\\
&= \rho\mathbf{v} \cdot \mathbf{g} + \nabla \cdot \left(\mathsf{K}\nabla e_\mathrm{cr}\right) + \dot{u}_\mathrm{chem} + \dot{u}_\mathrm{sn} + Q_\mathrm{cr}\\
\frac{\partial\mathbf{B}}{\partial t} &- \nabla \times \left(\mathbf{v} \times \mathbf{B}\right) = 0\\
\frac{\partial e_\mathrm{cr}}{\partial t} &+ \nabla \cdot \left(e_\mathrm{cr}\mathbf{v}\right) = -P_\mathrm{cr}\nabla\cdot\mathbf{v}+\nabla\cdot\left(\mathsf{K}\nabla e_\mathrm{cr}\right) + Q_\mathrm{cr},
\end{align}
with the mass density $\rho$, the gas velocity $\mathbf{v}$, the magnetic field $\mathbf{B}$, the total pressure $P_\mathrm{tot} = P_\mathrm{thermal} + P_\mathrm{magnetic} + P_\mathrm{cr}$, the total energy density $e = \frac{\rho v^2}{2} + e_\mathrm{thermal} + e_\mathrm{cr} + \frac{B^2}{8\pi}$, the momentum input of unresolved SNe $\dot{\mathbf{q}}_\mathrm{sn}$, the thermal energy input from resolved SNe, $\dot{u}_\mathrm{sn}$, the changes in thermal energy due to heating and cooling, $\dot{u}_\mathrm{chem}$, the CR diffusion tensor, $\mathsf{K}$, and the CR energy source term, ${Q_\mathrm{cr} = Q_\mathrm{cr, injection} + \Lambda_\mathrm{hadronic}}$. With each SN event, we inject energy into the CRs with the canonical amount of 10 per cent of the SN energy ($E_\mathrm{CR} = 10^{50}\,\mathrm{erg}$, see e.g. \citet{Ackermann2013}).
We assume a steady-state energy spectrum and choose a fixed diffusion coefficient along the magnetic field lines of $K_\parallel = 10^{28}\,\mathrm{cm^2\,s^{-1}}$ and of $K_\bot = 10^{26}\,\mathrm{cm^2\,s^{-1}}$ perpendicular to the magnetic field \citep{Strong2007, Nava2013}. 
CRs can cool through hadronic losses via pion decay after interacting inelastically with nuclei of the surrounding gas and through adiabatic expansion with an adiabatic index of $\gamma_\mathrm{CR} = \frac{4}{3}$. This leads to a general effective adiabatic index $\gamma_\mathrm{eff} = \frac{\gamma P_\mathrm{thermal} + \gamma_\mathrm{cr}P_\mathrm{cr}}{P_\mathrm{thermal} + P_\mathrm{cr}}$, with $\gamma = \frac{5}{3}$. We follow the prescription of \citet{Pfrommer2017} for the hadronic losses with
\begin{align}
\Lambda_\mathrm{cr} = - 7.44 \times 10^{-16} \left(\frac{n_\mathrm{e}}{\mathrm{cm^{-3}}}\right) \left(\frac{e_\mathrm{cr}}{\mathrm{erg\,cm^{-3}}}\right) \mathrm{erg\,s^{-1}\,cm^{-3}},
\end{align}
with $n_\mathrm{e}$ the free electron number density. We further discuss the parameter choices and limitations of our CR model in Sec. \ref{sec:crmodel}.

\subsection{Initial conditions}

The base setup models a galactic patch within a $500 \times 500 \times \pm 4000\,\mathrm{pc^3}$ computational domain with periodic boundary conditions along the $x-$ and $y-$ directions and strictly outflow boundary conditions along the extended $z-$direction. We enforce a fixed grid resolution of $dx \approx 3.9\,\mathrm{pc}$ within $|z| = 1\,\mathrm{kpc}$ and a base resolution of $dx \approx 7.8\,\mathrm{pc}$ beyond 1 kpc, which is allowed to refine up to $dx \approx 3.9\,\mathrm{pc}$ using the \textsc{flash} AMR architecture outside of that region. The magnetohydrodynamic equations are solved with a modified 3-wave Bouchut solver for the ideal MHD \citep{Bouchut2007, Waagan2011}, which includes cosmic ray pressure as an additional source term \citep{Girichidis2018, Girichidis2020}. The gas is initially setup with a Gaussian profile in pressure equilibrium with a uniform background density of $\rho_\mathrm{bg} = 10^{-27}\,\mathrm{g\,cm^{-3}}$. We vary the initial total gas surface density, $\Sigma_\mathrm{gas}$, and the standard deviation of the Gaussian profile, $\sigma_\mathrm{gas}$, between our models (see Table \ref{tab:runs}). To prevent the medium from collapsing into a thin uniform sheet under the influence of gravity, we artificially introduce turbulence by injecting kinetic energy on the largest scales in Fourier space to maintain a chosen root mean square velocity, $v_\mathrm{rms}$, up until the first sink particles form and star formation and stellar feedback take over the turbulent driving. We ensure that the injected turbulence consists of a mixture of 2:1 solenoidal to compressive modes \citep{Konstandin2015}.

\setlength{\tabcolsep}{4pt}
\begin{table}
	\centering
	\caption{List of simulations with their varying initial parameters. From left to right, we give the name of each simulation, the initial gas surface density $\Sigma_\mathrm{gas}$, whether or not CRs are included in the model, the constant strength of the far-ultraviolet interstellar radiation field, $G_0$, the initial strength of the magnetic field, $|\mathbf{B}|$, the thickness of the initial Gaussian gas density profile, $\sigma_\mathrm{gas}$, the target root mean square velocity of the initial turbulent driving, $v_\mathrm{rms}$, and lastly the total simulated time of the evolution since the onset of star formation, $t_\mathrm{evol}$. Models with the dagger symbol, $\dagger$, in their name do not include CRs. The test model $\Sigma100$-noFB has the same initial conditions and ISM processes included as $\Sigma100$ but turns off all stellar feedback processes (winds, ionising radiation, SNe, CR acceleration). See Appendix Fig. \ref{fig:noFB} for a discussion.}
	\begin{tabular}{lccccccc}
		\hline
		Name                      & $\Sigma_\mathrm{gas}$ & CR  & $G_0$ & |\textbf{B}|      & $\sigma_\mathrm{gas}$ & $v_\mathrm{rms}$ & $t_\mathrm{evol}$ \\
		                           & [M$_\odot$ pc$^{-2}$] &     &       & [$\mu\mathrm{G}$] &  [pc]                 & [km s$^{-1}$]    & [Myr]             \\
		\hline
		$\mathbf{\Sigma010}$      & 10                    & yes & 1.7   &  6                & 30                  & 10               & 250.4            \\
		$\mathbf{\Sigma030}$      & 30                    & yes & 7.9   & 10                & 37                  & 15               & 221.5            \\
		$\mathbf{\Sigma050}$      & 50                    & yes & 16.2  & 13                & 45                  & 20               & 203.7            \\
		$\mathbf{\Sigma100}$      & 100                   & yes & 42.7  & 19                & 60                  & 30               & 194.6            \\\\
		$\mathbf{\Sigma010^\dag}$ & 10                    & no  & 1.7   &  6                & 30                  & 10               & 273.9            \\
		$\mathbf{\Sigma100^\dag}$ & 100                   & no  & 42.7  & 19                & 60                  & 30               & 114.6            \\\\
		$\mathbf{\Sigma100}$\textbf{-noFB} & 100          & no  & 42.7  & 19                & 60                  & 30               & 117.5            \\
		\hline
	\end{tabular}
	\label{tab:runs}
\end{table}

In total, we perform a suite of seven simulations divided into a set of four simulations that include the nonthermal feedback channel of CRs and with varying initial surface density, $\Sigma_\mathrm{gas}$, labelled $\Sigma010$, $\Sigma030$, $\Sigma050$, $\Sigma100$, and another set of two simulations that omit CR injection and transport, $\Sigma010^\dagger$ and $\Sigma100^\dagger$. Finally, we simulate one more test model similar to $\Sigma100$ but with stellar feedback turned off. The label indicates the initial gas surface density $\Sigma_\mathrm{gas}$ ($\Sigma010 \rightarrow \Sigma_\mathrm{gas} = 10\,\mathrm{M_\odot\,pc^{-2}}$, etc.). See Table \ref{tab:runs} for a complete list of the seven simulations with their respective initial conditions. The goal is to model different galactic ISM environments by varying $\Sigma_\mathrm{gas}$, as well as to understand the impact of CRs, especially in high-density regimes, by turning this feedback channel on and off. Other parameters changed between the different models are the value of the constant far-ultraviolet (FUV) ISRF constant background $G_0$, the strength of the initial magnetic field $|\mathbf{B}|$, the thickness of the initial Gaussian gas density profile $\sigma_\mathrm{gas}$ and the strength of the initial turbulent driving with a target root mean square velocity, $v_\mathrm{rms}$. 
Please note that we do not change the metallicity or the external gravitational potential between our models.

\section{The simulated interstellar medium}\label{sec:TheSimulatedInterstellarMedium}

\begin{figure*}
	\centering
	\includegraphics[width=.529\linewidth]{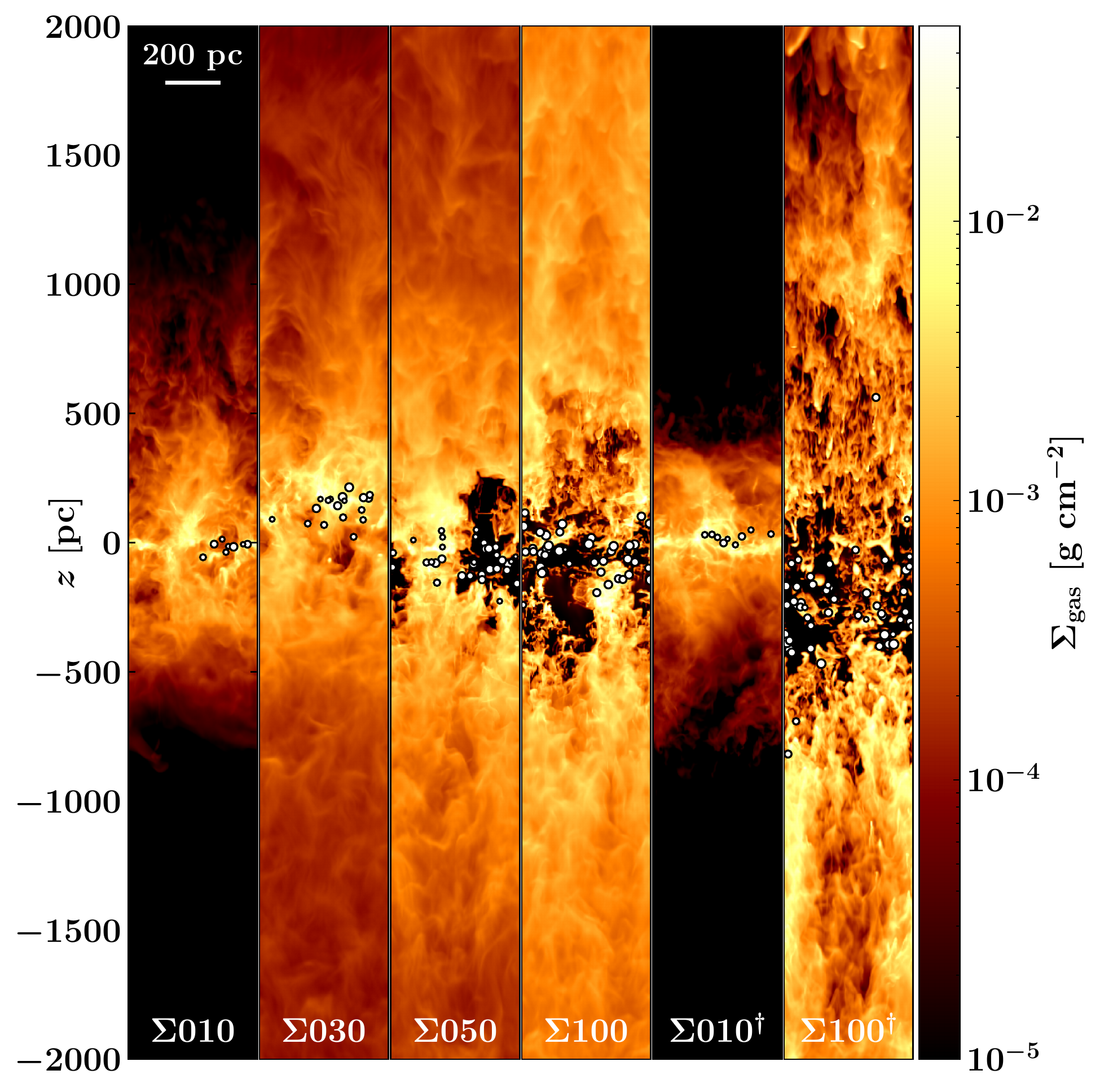}
	\includegraphics[width=.462\linewidth]{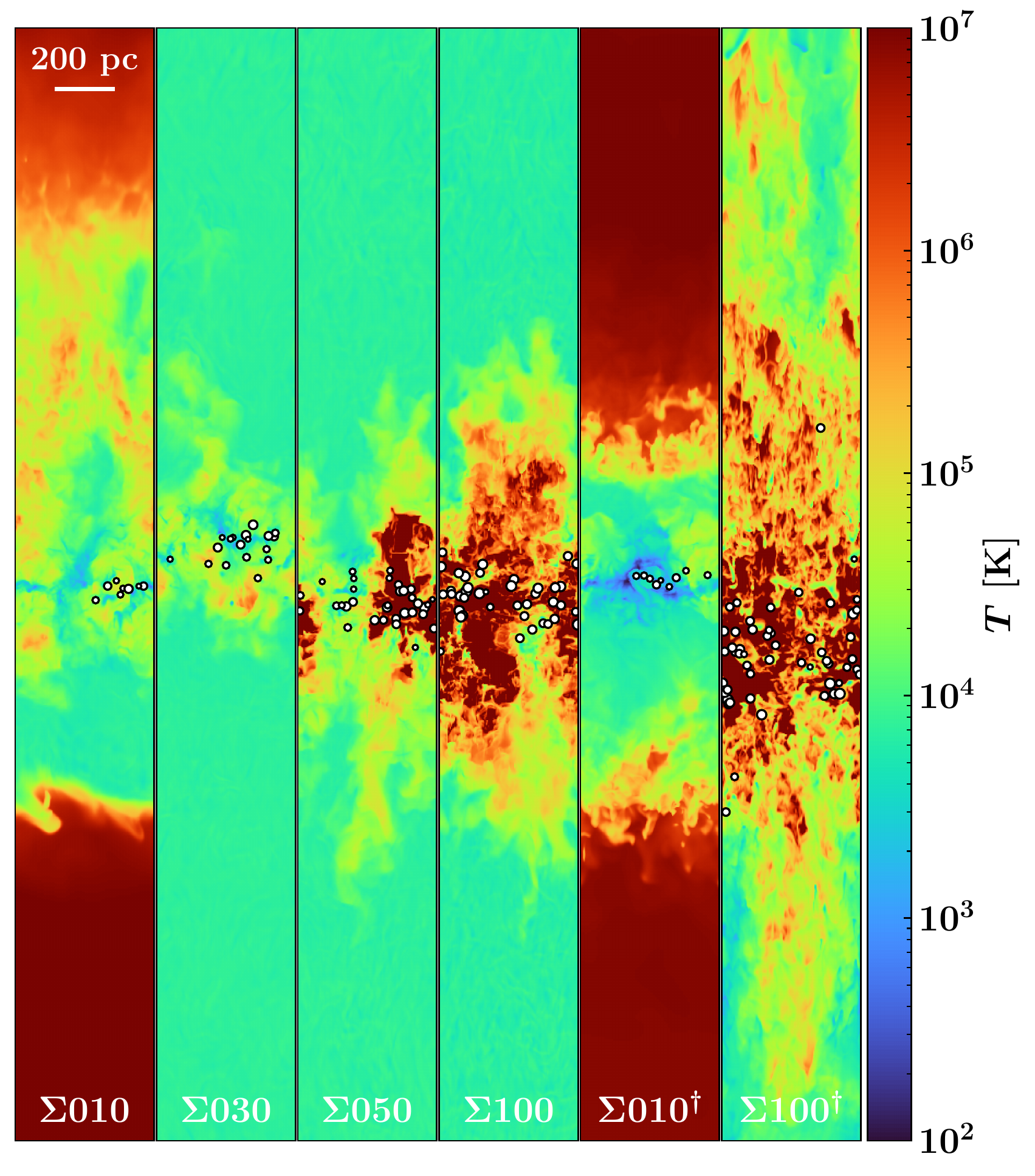}
	\caption{Edge-on views of the simulations at a characteristic snapshot $t - t_\mathrm{SFR} = 60\,\mathrm{Myr}$ with $t_\mathrm{SFR}$ being the onset of star formation after $\sim 20\,\mathrm{Myr}$ of evolution with artificial turbulent driving. We show the total gas column density $\Sigma_\mathrm{gas}$ (\textit{left panel}) and density-weighted temperature $T$ in projection (\textit{right panel}). The first four columns in each panel show the simulations with increasing initial gas surface density, ranging from $10\,\mathrm{M}_\odot\,\mathrm{pc}^{-2}$ to $100\,\mathrm{M}_\odot\,\mathrm{pc}^{-2}$. The last two columns ($\Sigma010^\dag$ \& $\Sigma100^\dag$) show runs with the lowest and highest gas surface density initial conditions but without the inclusion of CRs. We only show the simulated volume up to a height of $|z| = 2\,\mathrm{kpc}$, while the real simulated volume spans up to $|z| = 4\,\mathrm{kpc}$. White circles represent active star clusters with their drawn sizes scaled by the cluster's mass. Higher initial gas surface densities promote stronger star formation and drive stronger outflows. The omission of CRs at the higher surface density leads to a more structured, patchier and hotter outflow (compare column $\Sigma100$ and $\Sigma100^\dag$ in the right panel). CRs also drive an outflow in lower-density environments (compare $\Sigma010$ and $\Sigma010^\dag$ in the left panel).}
	\label{fig:overview}
\end{figure*}

To introduce our suite of simulations, we show a general overview of the simulation morphologies in Fig. \ref{fig:overview}. Additionally, the reader can find an animation of the evolution of the ISM for model $\Sigma100$ on the \textsc{silcc} homepage\footnote{\url{https://hera.ph1.uni-koeln.de/~silcc/\#downloads}}. In the left panels of Fig. \ref{fig:overview}, we show the gas column densities, $\Sigma_\mathrm{gas}$, and in the right panels, the density-weighted temperature projection, $T$, both viewed edge-on. The first four columns in each panel show from left to right the setups with an increase in the initial gas surface density from $\Sigma_\mathrm{gas} = 10\,\mathrm{to\,}100\,\mathrm{M_\odot\,pc^{-2}}$. The second to last and right columns show the models without the inclusion of CRs for the lowest and highest initial surface density, $\Sigma010^\dagger$ and $\Sigma100^\dagger$, respectively. Star cluster sink particles are visualised as white circles. The size of the circles scales with the mass of the star cluster and does not represent the actual physical size of an individual cluster\footnote{please note that we do not resolve internal dynamics within a star cluster sink particle. Regarding sink-sink interactions, the numerical size of all our sink particles is equal to their fixed accretion radius, that is, $\sim12\,\mathrm{pc}$, as described in Section \ref{sec:numerics}. The gravity module treats the sink particles as point masses.}. All snapshots are at $t - t_\mathrm{SFR} = 60\,\mathrm{Myr}$ with $t_\mathrm{SFR}$ being the onset of star formation after $\sim 20\,\mathrm{Myr}$, which we have chosen as a characteristic representation of the simulations.
Higher initial surface densities result in stronger star formation and promote more extensive outflows. Although the gas barely reaches a height of $z = 1\,\mathrm{kpc}$ for $\Sigma010$ after $60\,\mathrm{Myr}$, it already passes through the $z=2\,\mathrm{kpc}$ boundary for $\Sigma030$ and beyond. At the high end of the initial conditions, $\Sigma050$ and $\Sigma100$, star formation becomes so violent that it starts to tear holes in the midplane ISM and fully depletes the gas reservoir needed to form new stars. However, the gas structure in the outflow regions farther away from the midplane appears smooth and unstructured. This picture changes for models that omit CR injection and acceleration ($\Sigma010^\dagger$, $\Sigma100^\dagger$). For the lowest surface density model, the CRs do not seem to have a strong effect on the midplane ISM, however, there is no outflow present without CRs in $\Sigma010^\dagger$ 60 Myr after the onset of star formation. The picture changes drastically for the highest surface density run, $\Sigma100^\dagger$. The midplane ISM seems even more perforated by hot superbubbles and the morphology of the outflow is much more patchy and structured than the corresponding simulations without CRs \citep[see e.g.][]{Girichidis2018}. A more detailed analysis of the outflow will follow in Section \ref{sec:StarFormationAndOutflows}.
The qualitative change of the ISM and the outflow structure based on the gas column densities is also reflected in the mass-weighted temperature projections (right panels of Fig. \ref{fig:overview}). Stars are formed in pockets of cold gas (seen in blue). With increasing gas surface density and hence increasing star formation activity, overlapping bubbles created by stellar winds and SNe are formed, which create a large volume filling of the hot gas phase (seen in red). This evacuates the midplane ISM from the star-forming gas reservoir and drives outflows. However, the outflows themselves cool down, and the outflow gas appears to exist mainly in the warm gas phase at temperatures around $T\approx10^4\,\mathrm{K}$. When the nonthermal CR feedback is left out, the phase structure of the outflow changes and patches of hot gas percolate through medium to large heights. Note that the large volumes of $T > 10^7\,\mathrm{K}$ gas in $\Sigma010$ and $\Sigma010^\dagger$ above and below the midplane result from the numerical initial conditions of the simulations. This is not a gas that got transported to those heights from the midplane, but a low-density halo gas of our stratified disk model, which is set to high temperatures to be in initial pressure equilibrium.

\begin{figure}
	\centering
	\includegraphics[width=.99\linewidth]{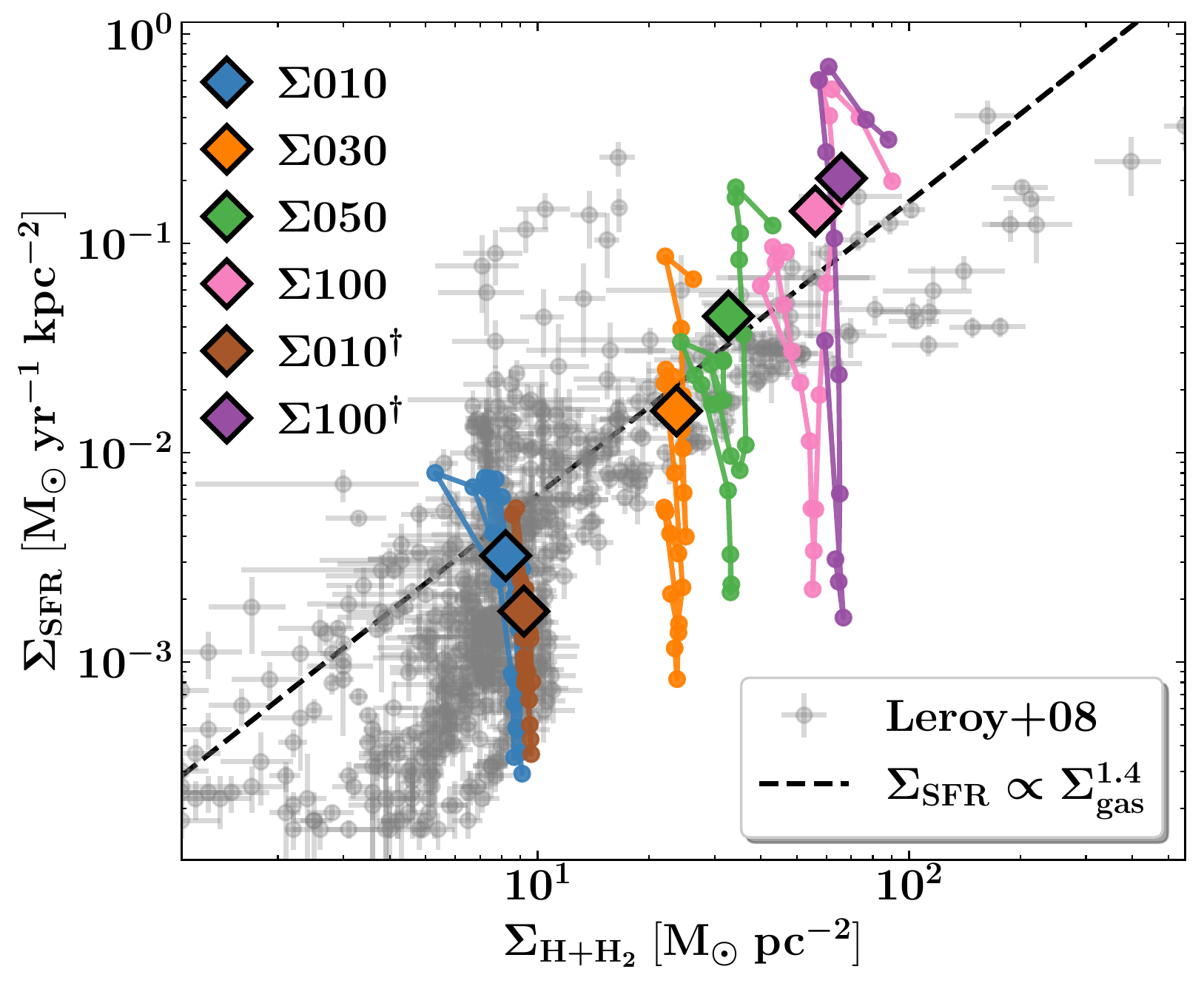}
	\caption{Average star formation rate surface density $\Sigma_\mathrm{SFR}$ as a function of the average gas surface density $\Sigma_\mathrm{H+H_2}$. The diamonds represent global averages whereas the connected circles are subsequent 10 Myr bins. Shown in grey are observational data of nearby resolved galactic patches by \citet{Leroy2008}. The dashed black line indicates the Kennicutt-Schmidt relation \citep{KennicuttJr.1998} centred on the observed mean star formation rate surface density at $\Sigma_\mathrm{H+H_2} = 10\,\mathrm{M_\odot\,pc^{-2}}$. Especially the higher surface density models undergo a phase of high initial star formation followed by a drop in star formation rate, before self-regulation kicks in. On average, our models, which have no in-built scaling with gas density, are close to observational values.}
	\label{fig:kennicutt}
\end{figure}

We give a more quantitative overview of the evolution of our ISM simulations in the Kennicutt-Schmidt-relation \citep{KennicuttJr.1998} plot, Fig. \ref{fig:kennicutt}. We show the average star formation rate surface density, $\Sigma_\mathrm{SFR}$, as a function of the average atomic and molecular gas surface density, $\Sigma_\mathrm{H+H_2}$ for $10\,\mathrm{Myr}$ bins beginning with the onset of star formation, $t_\mathrm{SFR}$ (small circles in Fig. \ref{fig:kennicutt}). The successive time bins are connected by a solid line. The larger diamonds indicate the global averages until the end of the simulation (see Table \ref{tab:runs}). We also show observational data from resolved galactic patches of nearby star-forming galaxies compiled by \citet{Leroy2008}. The dashed black line indicates the Kennicutt-Schmidt relation $\Sigma_\mathrm{SFR} \propto \Sigma_\mathrm{gas}^{1.4}$. We anchor the line on the average $\Sigma_\mathrm{SFR}$ in the $\Sigma_\mathrm{H+H_2} = 10\,\mathrm{M_\odot\,pc^{-2}}$ bin of the observational data. However, this choice is not unambiguous since there is a large scatter in the observational data, especially for solar neighbourhood conditions.
Initially, our models undergo a strong period of star formation or even a starburst for the higher surface densities ($\Sigma030$, orange; $\Sigma050$, green; $\Sigma100$, pink; $\Sigma100^\dagger$). This initial starburst phase is influenced by the initial condition of our simulation setup. A result of this elevated star formation is the fast depletion of star-forming gas, which leads to a steep drop-off in the star formation rate. After this first evolutionary phase, the now preprocessed ISM undergoes subsequent episodes of star formation and levels out at star formation rates, which are in good agreement with observational estimates.

\section{Star formation and outflows}\label{sec:StarFormationAndOutflows}

In this section, we give a more detailed analysis of the star formation characteristics and outflow properties. We focus on the mass and energy loading factors and the multiphase nature of the outflow, which is strongly influenced by the presence of CRs. We define the mass loading factor, $\eta$, as the ratio of $\dot{M}_\mathrm{out}$, the instantaneous mass outflow rate through a boundary (either $|z| = 1\,\mathrm{or\,}2\,\mathrm{kpc}$), and ${<\mathrm{SFR}>}$, the global average star formation rate,
\begin{equation}\eta = \dot{M}_\mathrm{out} / <\mathrm{SFR}>.\end{equation}
Similarly, we define the energy loading factor, $\gamma$, as the ratio of $\dot{E}_\mathrm{out}$, the total instantaneous energy outflow rate, including terms for the thermal, kinetic, magnetic, and CR energy, and $<\dot{E}_\mathrm{SN}>$, the globally averaged SN injection energy. These are the same definitions as used in the previous work \citet{Rathjen2021}. The quantitative results of this section are summarised in Table \ref{tab:sfr_of_global}.

\begin{figure*}
	\centering
	\includegraphics[width=.515\linewidth]{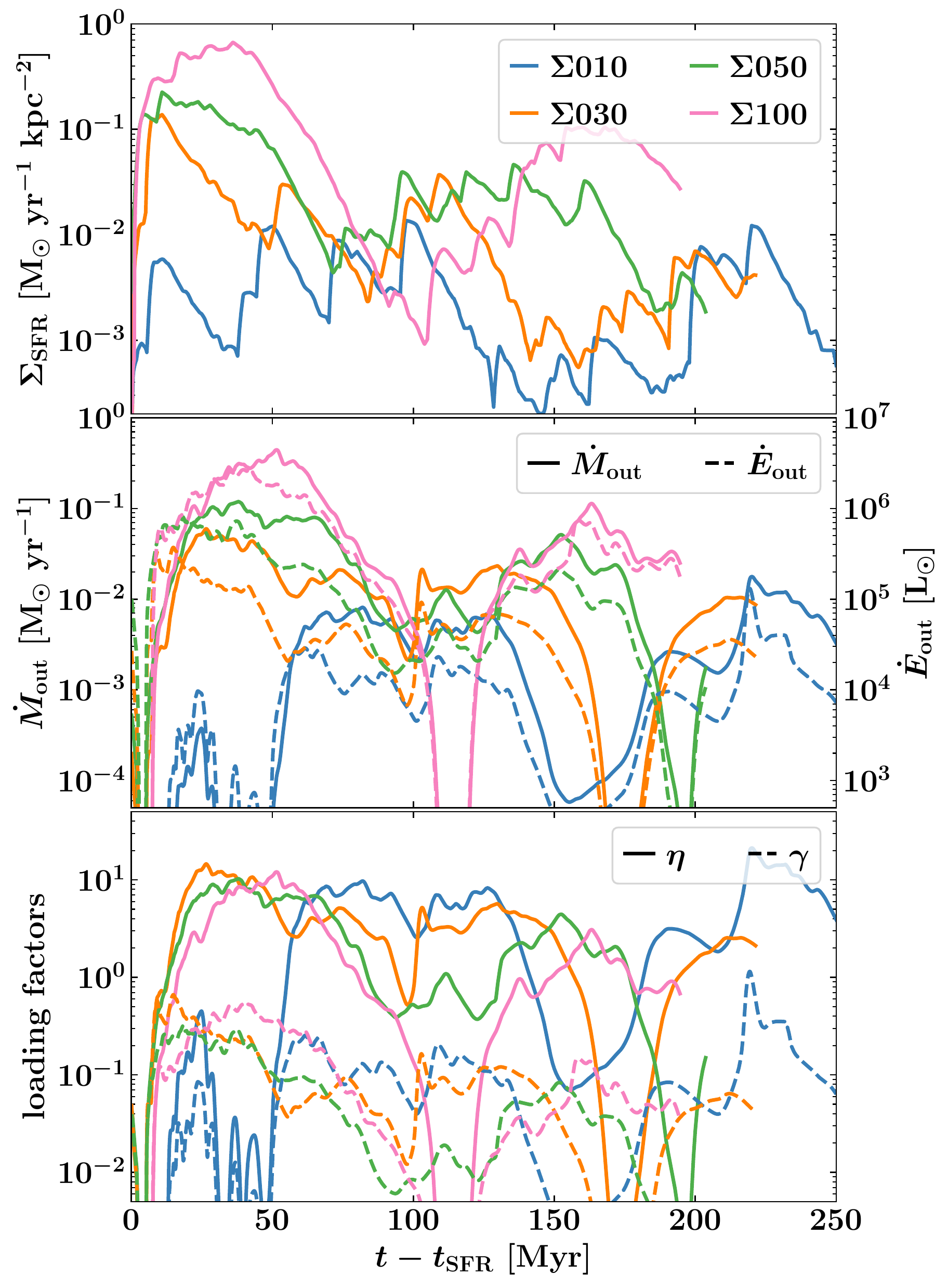}
	\includegraphics[width=.475\linewidth]{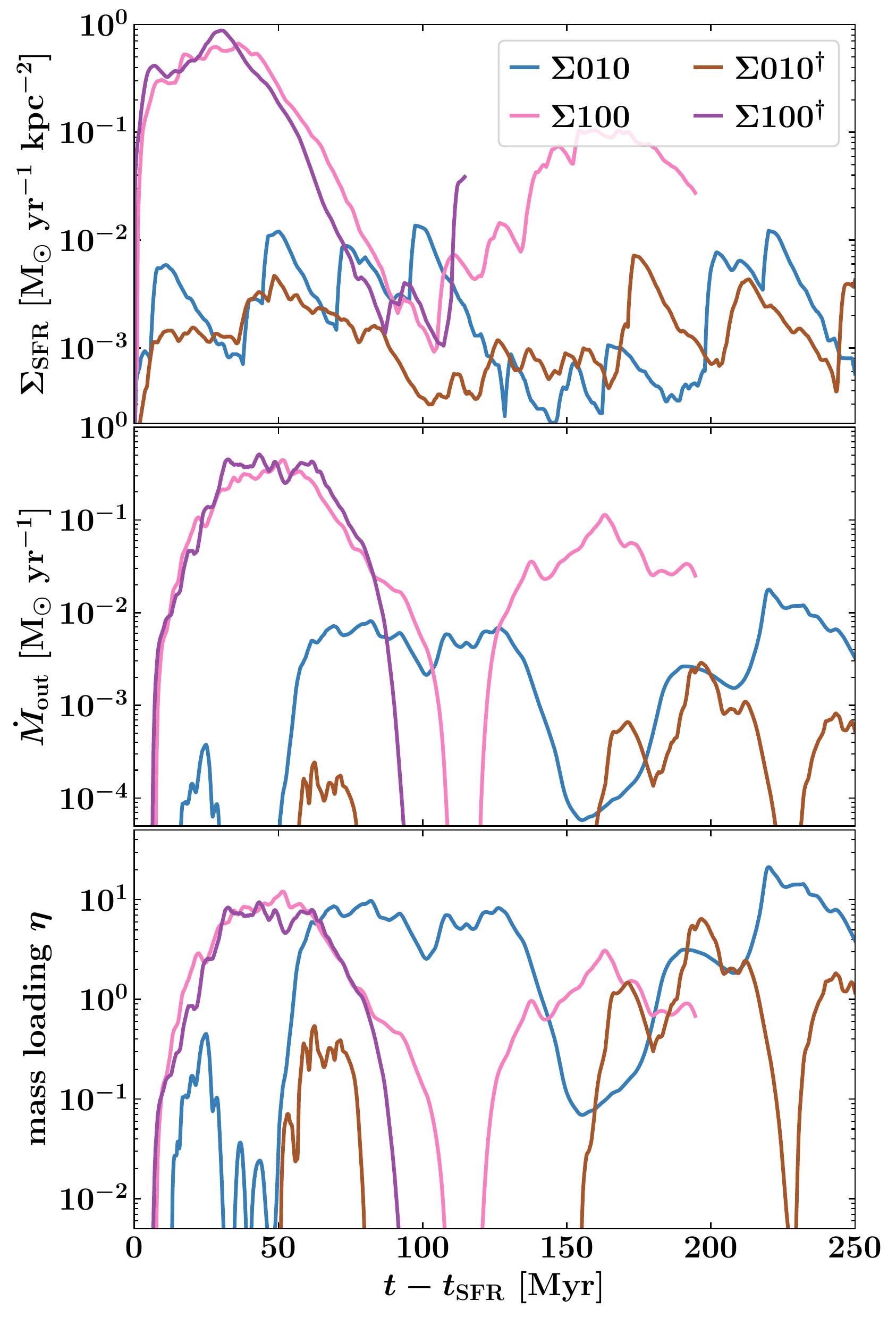}
	\caption{\textit{Left panels}: Star formation rate surface density, $\Sigma_\mathrm{SFR}$ (top panel), mass outflow rate, $\dot{M}_\mathrm{out}$, and energy outflow rate, $\dot{E}_\mathrm{out}$, through $|z| = 1\,\mathrm{kpc}$, (middle panel), and mass loading factor, $\eta = \dot{M}_\mathrm{out} / <\mathrm{SFR}>$, and energy loading factor, $\gamma = \dot{E}_\mathrm{out} / <\dot{E}_\mathrm{SN}>$, (bottom panel), for the four simulations including CRs. Here and in all following plots, we show the evolution beginning with the onset of star formation, respectively. Runs with a higher initial gas surface density experience a larger starburst which results in an up to two orders of magnitude higher mass outflow rate than in $\Sigma010$. The outflow rate in $\Sigma010$ picks up towards the end of the simulated evolution because of the dynamical impact of cosmic rays, which take a longer time scale $\Delta t \approx 50-75\,\mathrm{Myr}$ to become effective \citep[see e.g. ][]{Girichidis2018, Girichidis2022}. The strong outflow leads to gas depletion in the midplane ISM, reducing the overall star formation rate. This allows gas to accumulate again in the midplane and star formation starts again.\\
		\textit{Right panels}: Comparison to models without CRs. Note that we focus only on mass outflow and mass loading factors and omit the energy equivalents. In low-density environments, CRs have a strong impact on mass outflow properties. Even though star formation rates are comparable, the mass outflow rate with CRs is boosted by up to two orders of magnitude during periods of low star formation. For the high surface density environment, CRs do not seem to change global values of star formation and mass outflow rates (see however Fig. \ref{fig:vouts_010} and Appendix Fig. \ref{fig:vouts_100} on the different structure of the outflow). They can cool more efficiently via hadronic losses in the more dense medium and are less able to establish the additional pressure gradient. However, due to computational cost, we could only let $\Sigma100^\dag$ evolve for $\sim 100\,\mathrm{Myr}$ after the onset of star formation.}
	\label{fig:sfr_eta_gamma}
\end{figure*}

\begin{figure*}
	\centering
	\includegraphics[width=.515\linewidth]{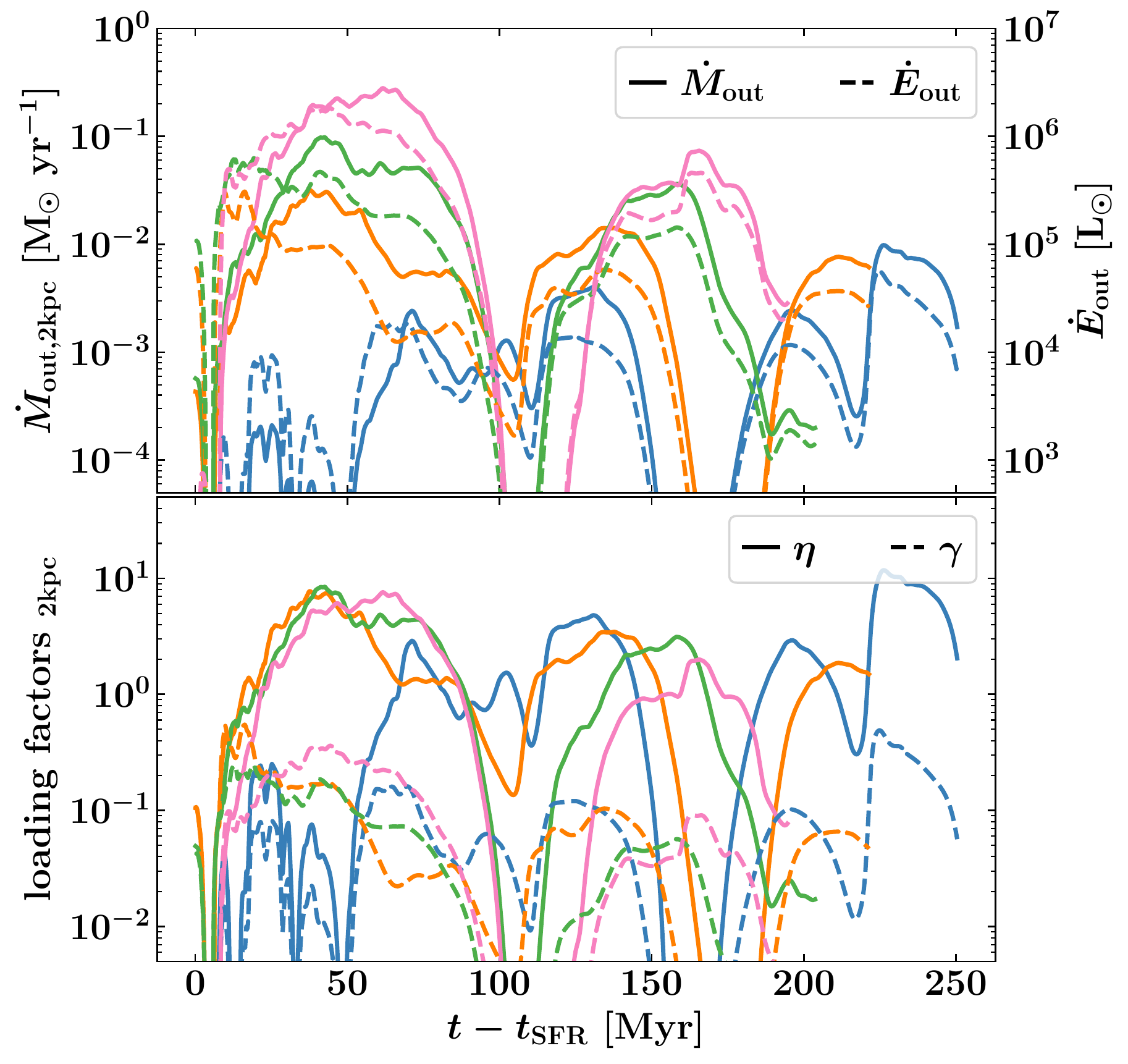}
	\includegraphics[width=.475\linewidth]{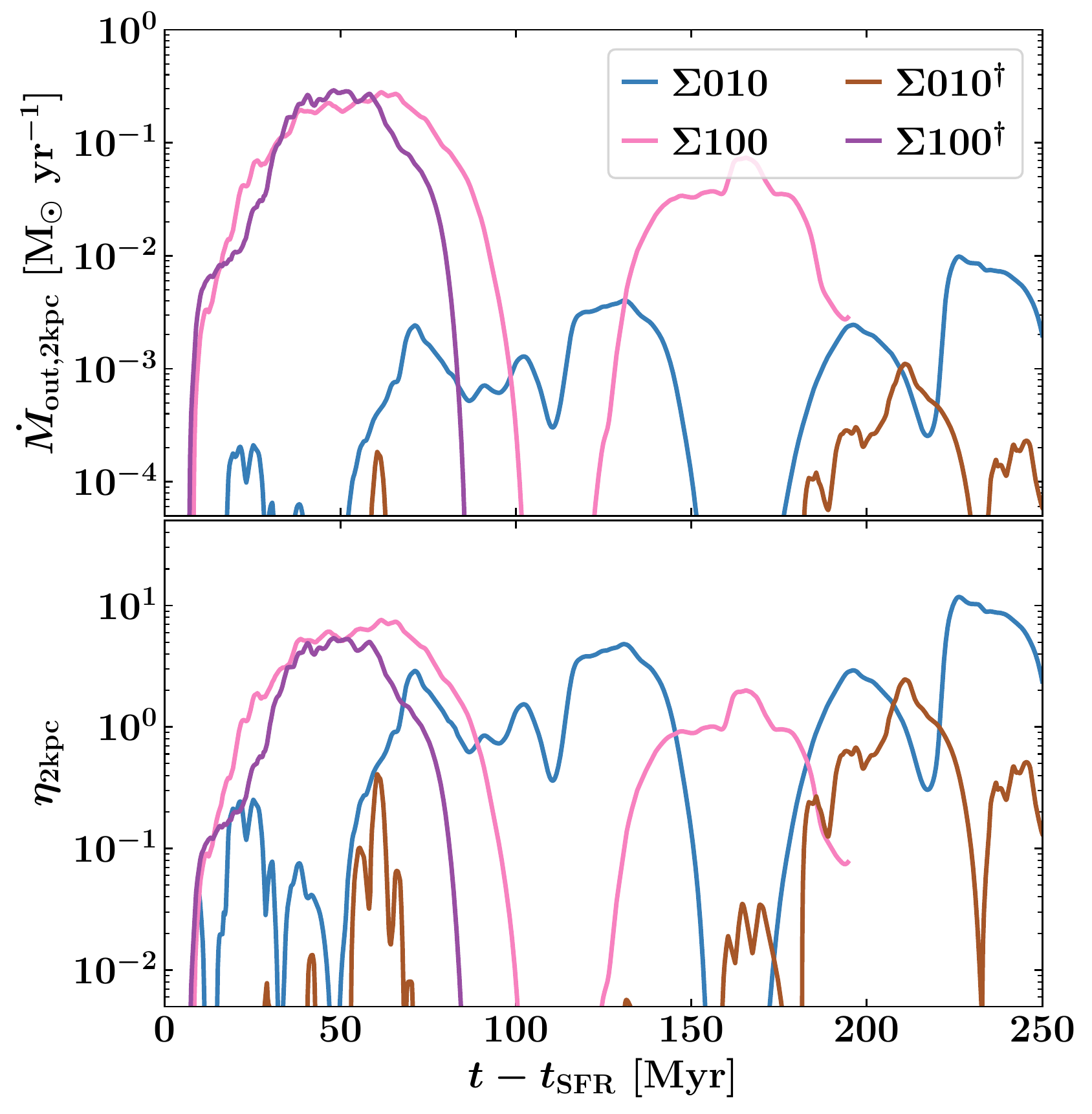}
	\caption{Same as Fig. \ref{fig:sfr_eta_gamma} but with outflow properties measured at $|z| = 2\,\mathrm{kpc}$ instead of $|z| = 1\,\mathrm{kpc}$. The lower initial surface density runs experience a stronger drop-off in mass loading $\eta$ while $\Sigma100$ shows the same median mass loading at the two measured heights ($|z| = 1\,\mathrm{and\,}2\,\mathrm{kpc}$) of $\eta_\mathrm{1kpc} = \eta_\mathrm{2kpc} = 0.9$ (see Table \ref{tab:sfr_of_global}). It is evident that CRs enable a high mass loading outflow for solar neighbourhood conditions in periods of low star formation (blue and brown lines in the bottom panel on the right), whilst the high-density environments do not show differences in the total transported mass out of the midplane ISM (pink and purple lines in the bottom panel on the right).}
	\label{fig:loading2kpc}
\end{figure*}

We analyse the star formation rate surface density, $\Sigma_\mathrm{SFR}$, total mass outflow rate, $\dot{M}_\mathrm{out}$, and the energy and mass loading factors, $\gamma$ and $\eta$, in Fig. \ref{fig:sfr_eta_gamma}. On the left-hand side of the figure, we focus on the 4 models including CR injection and acceleration, whereas on the right-hand side we show the direct comparison of the simulations lacking this nonthermal feedback channel. All outflow rates and loading factors shown here are calculated at a boundary of $|z| = 1\,\mathrm{kpc}$. The same analysis, measured at a height of $|z| = 2\,\mathrm{kpc}$ is presented in Fig. \ref{fig:loading2kpc}.
The star formation in our models is cyclical, best seen in the evolution of $\Sigma010$ (blue line, top left panel of Fig. \ref{fig:sfr_eta_gamma}). $\Sigma_\mathrm{SFR}$ first rises to a local maximum. Stellar feedback and further depletion of gas stop subsequent star formation, and $\Sigma_\mathrm{SFR}$ slowly declines again. However, this enables gas to accumulate again in the midplane, and a new episode of rapid star formation commences. This leads to a typical sawtooth shape of the $\Sigma_\mathrm{SFR}$ curve. The time frame of one star formation cycle is of the order of $20-50\,\mathrm{Myr}$. We calculate a star formation efficiency, SFE, as the ratio of the gas mass transformed into new stars to the total available gas mass within the sink particle accretion radius and achieve an SFE between $\mathrm{SFE} = 2.3 - 4.9\,\mathrm{per\,cent}$ in all of our models. The behaviour of the higher surface density models ($\Sigma030$: orange; $\Sigma050$: green; $\Sigma100$: pink) is similar, but with more pronounced features. The initial starburst is stronger for higher surface densities and reaches star formation rate surface densities of up to $\Sigma_\mathrm{SFR} \approx 1\,\mathrm{M_\odot\,yr^{-1}\,kpc^{-2}}$ for $\Sigma100$. Also, the time for the first starburst to subside increases. It takes $\sim100\,\mathrm{Myr}$ in the case of $\Sigma100$ to reach the first local minimum in star formation after the initial starburst as compared to $\sim40\,\mathrm{Myr}$ in the case of solar neighbourhood conditions, $\Sigma010$. We calculate a characteristic $\Sigma_\mathrm{SFR}$ for each simulation as the median of $\Sigma_\mathrm{SFR}$ since the beginning of star formation, $t_\mathrm{SFR}$, until the end of the simulation, $t_\mathrm{end}$, which extends over two orders of magnitude between $\Sigma010$ with $\Sigma_\mathrm{SFR} = 3 \times 10^{-3}\,\mathrm{M_\odot\,yr^{-1}\,kpc^{-2}}$ and $\Sigma100$ with $\Sigma_\mathrm{SFR} = 1.46 \times 10^{-1}\,\mathrm{M_\odot\,yr^{-1}\,kpc^{-2}}$. All values are also tabulated with their \nth{25} and \nth{75} percentiles as lower and upper bounds in Table \ref{tab:sfr_of_global}.

We show the total mass and energy outflow rates through $|z| = 1\,\mathrm{kpc}$, $\dot{M}_\mathrm{out}$ and $\dot{E}_\mathrm{out}$, in the middle panel on the left side of ${\mathrm{Fig.\,\ref{fig:sfr_eta_gamma}}}$. Solid lines represent $\dot{M}_\mathrm{out}$ (left y-axis) and dashed lines $\dot{E}_\mathrm{out}$ (right y-axis). Qualitatively, the two respective curves of each model proceed similarly with their peak values scaling with the initial gas surface density. Within the first $50\,\mathrm{Myr}$, $\Sigma010$ experiences a short period of outflow which is driven by the first episode of star formation. However, since star formation in $\Sigma010$ is generally moderate, this outflow does not carry much mass or energy and decays again. After $50\,\mathrm{Myr}$, stronger and longer-lasting mass and energy outflows are established, supported by the additional CR pressure gradient. This additional outflow driving agent needs to build up first and starts becoming dynamically important on a longer time scale. Once the CR pressure gradient is established, it is a long-lasting reservoir, since the CRs in our implementation only cool via adiabatic expansion and hadronic losses (see Section \ref{sec:numerics}). This is especially important in low-density environments such as $\Sigma010$, where hadronic losses are almost negligible. For the higher gas surface density runs, the outflows begin more immediately, only with a short delay after the onset of star formation. There exist short episodes ($\sim10\,\mathrm{Myr}$) in which the outflows completely cease due to the depletion of star-forming gas, either through feedback-triggered outflow or because it got locked up in newly formed stars (e.g. at $t - t_\mathrm{SFR}\approx100\,\mathrm{Myr}$ for $\Sigma100$).

In the bottom panel of the left-hand side of Fig. \ref{fig:sfr_eta_gamma}, we show the mass- and energy-loading factors, $\eta$ and $\gamma$. Again, the quantity associated with the mass outflow ($\eta$) is depicted as solid lines, whereas $\gamma$ is shown as dashed lines. In general, the characteristic (i.e. the median) mass loading, $\eta_\mathrm{1kpc}$, decreases with increasing initial $\Sigma_\mathrm{gas}$ from $\eta_\mathrm{1kpc} = 2.8$ for $\Sigma010$ to $\eta_\mathrm{1kpc} = 0.9$ for $\Sigma100$. A similar trend in energy loading, $\gamma$, is not detectable with $\gamma_\mathrm{1kpc} = (13, 12, 7, 10)\,\mathrm{per\,cent}$ for ($\Sigma010$, $\Sigma030$, $\Sigma050$, $\Sigma100$), respectively. Interestingly, all four $\Sigma_\mathrm{gas}$ models exhibit similar peak mass loading factors $\eta_\mathrm{1kpc} \approx 10$, independent of the initial density of the gas surface. A possible explanation for this behaviour is the dominant impact of the CR pressure gradient on driving cold and warm gas outflows, especially for low gas surface density systems. For the higher surface density systems $\Sigma030$, $\Sigma050$, $\Sigma100$, the first local maximum in $\eta_\mathrm{1kpc}$ is reached during the first starburst, but then the mass outflows slightly decrease over time again due to the decrease of star formation caused by direct stellar feedback and gas depletion in the midplane. However, in later stages the outflow rates increase again to a second local maximum, while the star formation rates only increase moderately, resulting in a mass loading at the same level as from the initial starburst. This happens around $t - t_\mathrm{SFR} \sim 100\,\mathrm{Myr}$ when the CRs have to build up the additional pressure gradient and become dynamically important, if not dominant. In the solar-neighbourhood condition simulation, $\Sigma010$, this effect is even clearer. The onset of star formation is smoother and less bursty, and there is no initial burst in the mass outflow. However, once the CRs have built up, again at about $t - t_\mathrm{SFR} \sim 50\,\mathrm{Myr}$, the mass outflow increases dramatically, leading to mass loading factors of $\sim 10$ and higher. 

On the right-hand side of Fig. \ref{fig:sfr_eta_gamma}, we present a direct comparison of the non-CR runs with their CRs including counterparts ($\Sigma010$, blue lines and $\Sigma010^\dagger$, brown lines; $\Sigma100$, pink lines and $\Sigma100^\dagger$, purple lines). Even though temporal fluctuations exist, $\Sigma_\mathrm{SFR}$ behaves alike between non-CR and CR models. The median star formation rate surface density (with \nth{25} and \nth{75} percentiles as lower and upper bounds) only slightly decreases from $\Sigma_\mathrm{SFR} = 3\times10^{-3}\,\mathrm{M_\odot\,yr^{-1}\,kpc^{-2}}$ in $\Sigma010$ to $\Sigma_\mathrm{SFR} = 2\times10^{-3}\,\mathrm{M_\odot\,yr^{-1}\,kpc^{-2}}$ in $\Sigma010^\dagger$. However, CRs have a strong impact in low-density environments in terms of outflows. CR-supported outflows are driven even in periods of low star formation rates, whereas without CRs, an outflow can only be established in the presence of a volume-filling hot gas phase generated by strong star formation and hence stellar feedback activity. This results in a low $\eta_\mathrm{1kpc} = 0.03$ for $\Sigma010^\dagger$ compared to $\eta_\mathrm{1kpc} = 2.8$ for its counterpart $\Sigma010$. In high-density environments, CRs have a lower impact on galactic outflows. As a result of the larger hadronic losses, CRs cool more efficiently and the additional long-lived CR pressure gradient is smaller in magnitude. Qualitatively, the mass outflow rate, $\dot{M}_\mathrm{out}$, and the mass loading, $\eta$, are not significantly affected by the inclusion of CR. However, we want to note that these are only zeroth-order results since the total time evolution of $\Sigma100^\dagger$ is much shorter than in the other models due to computational cost. Therefore, the quantitative results for $\Sigma100^\dagger$ must be taken with a grain of salt and are not based on solid statistical bases.

We repeat the analysis of Fig. \ref{fig:sfr_eta_gamma} in Fig. \ref{fig:loading2kpc} but measure the outflow properties at higher altitude, $|z| = 2\,\mathrm{kpc}$. The median energy loading factor drops only slightly between $|z| = 1\,\mathrm{to\,}2\,\mathrm{kpc}$ with $\gamma_\mathrm{2kpc} = (9, 9, 6, 8)\,\mathrm{per\,cent}$. Most of the energy transported away from the midplane ISM through $|z| = 2\,\mathrm{kpc}$ is carried by the CRs with $\dot{E}^\mathrm{out}_\mathrm{CR} / \dot{E}^\mathrm{out}_\mathrm{tot} = (73 \pm 8)\,\mathrm{per\,cent}$ across the four different initial $\Sigma_\mathrm{gas}$ models. The hadronic cooling losses of the CRs are small and the energy transport away from the midplane ISM is persistent in space and time once the long-lived CR pressure gradient is established over kpc scales. On the other hand, the decrease in the mass loading factors between $|z| = 1\,\mathrm{and\,}2\,\mathrm{kpc}$ is steeper, although it depends on the initial surface density. For the runs $\Sigma010$, $\Sigma030$, and $\Sigma050$ the mass loading factor is reduced from $\eta_\mathrm{1kpc} = (2.8, 2.8, 1.8)$ to $\eta_\mathrm{2kpc} = (0.8, 1.3, 1.0)$, respectively. For $\Sigma100$, the characteristic mass loading factor remains nearly constant between the two altitudes at $\eta = 0.9$ and we do not see a decrease in the total mass outflow between the two boundaries. The outflow is fuelled by the two strong starburst events (i.e. the initial starburst and the second episode of strong star formation starting at $t - t_\mathrm{SFR} \sim 100\,\mathrm{Myr}$) and gas efficiently driven up to heights exceeding $2\,\mathrm{kpc}$ by the hot gas phase created in overlapping SNR whilst additionally supported by CRs.
Without CRs, the low-mass system, $\Sigma010^\dagger$, lacks an outflow at $|z| = 2\,\mathrm{kpc}$ for most of the simulated time with active star formation. However, a short episode of a ballistic outflow launched out of the hot gas phase is established toward the end of the simulation ($t - t_\mathrm{SFR} \approx 200\,\mathrm{Myr}$) with a peak mass loading factor of the order of unity.

\begin{figure}
	\centering
	\includegraphics[width=.99\linewidth]{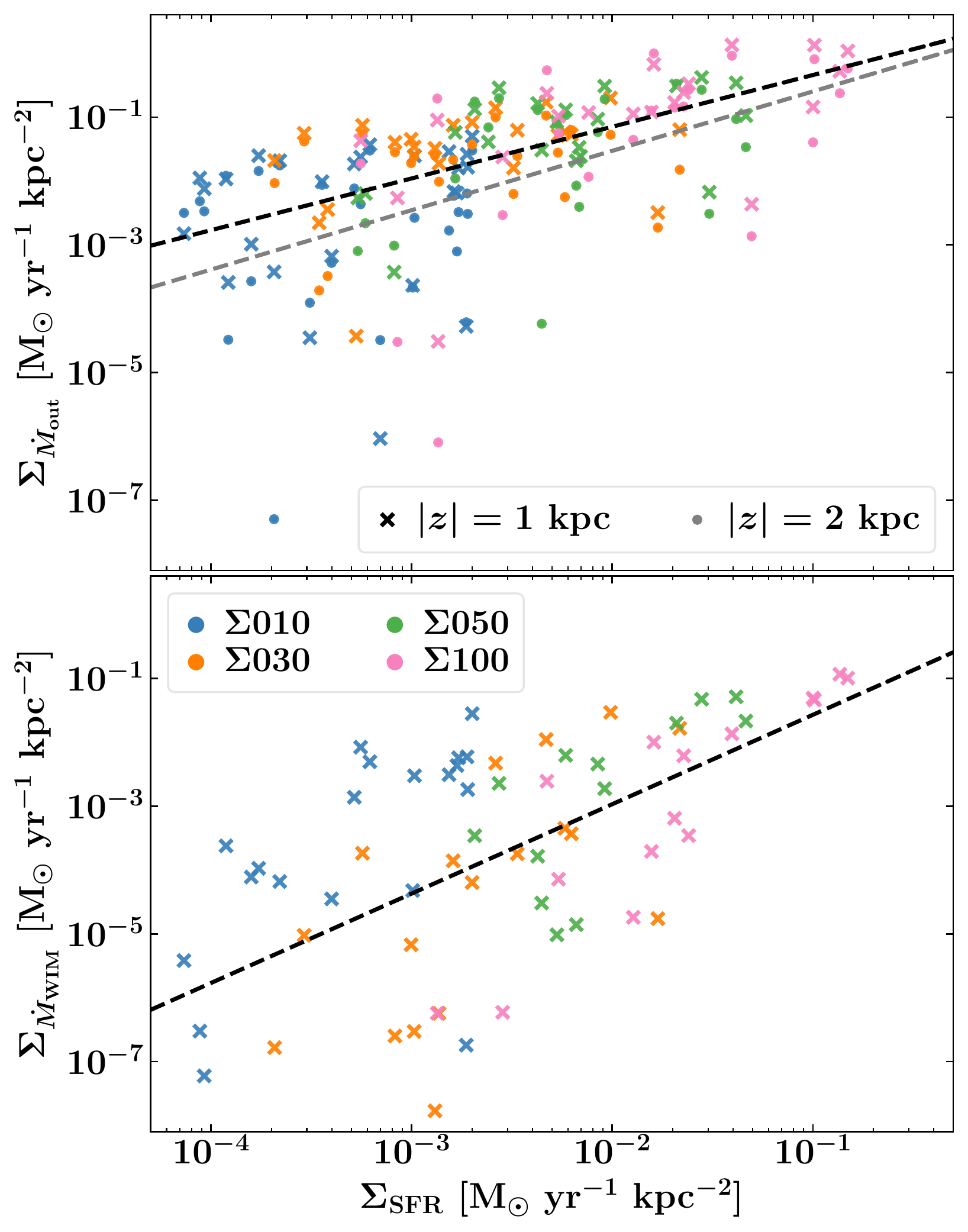}
	\caption{\textit{Top} Total gas mass outflow rate surface density, $\Sigma_{\dot{M}_\mathrm{out}}$, as a function of $\Sigma_\mathrm{SFR}$ measured at two different heights (crosses: 1 kpc, dots: 2 kpc). The dashed lines are power law fits (black: 1 kpc, grey: 2 kpc) with slopes $\alpha_\mathrm{1kpc}=0.81\pm0.12$ and $\alpha_\mathrm{2kpc}=0.93\pm0.18$. The data points are successively averaged values over $10\,\mathrm{Myr}$. The strength of the galactic outflow scales with star formation for all initial gas surface densities.
		\textit{Bottom} Same as the top but for the warm and ionised gas outflow rate surface density, $\Sigma_{\dot{M}_\mathrm{WIM}}$, through $|z| = 1\,\mathrm{kpc}$ as a function of $\Sigma_\mathrm{SFR}$. The dashed line is a power law fit with slope $\alpha = 1.40 \pm 0.24$.}
	\label{fig:of_sf}
\end{figure}

We show the surface density of the total mass outflow rate, $\Sigma_{\dot{M}_\mathrm{out}}$ through the two boundaries $|z| = 1\,\mathrm{and\,}2\,\mathrm{kpc}$ as a function of $\Sigma_\mathrm{SFR}$ in the top panel of Fig. \ref{fig:of_sf} for simulations that include CRs. The crosses indicate the outflow rate through $|z| = 1\,\mathrm{kpc}$ while dots represent the outflow rate through $|z| = 2\,\mathrm{kpc}$. Data points are successively averaged values over $10\,\mathrm{Myr}$. The power-law fits of the data are shown as dashed lines and have slopes of $\alpha_\mathrm{1kpc}=0.81\pm0.12$ and $\alpha_\mathrm{2kpc}=0.93\pm0.18$. The strength of the outflow is correlated with the star formation activity within the midplane ISM, although the correlation weakens between the two boundaries. The Spearman rank correlation coefficients, $R$, are $R_\mathrm{1kpc} = 0.69$ and $R_\mathrm{2kpc} = 0.58$, both with $p-\mathrm{values} \leq 10^{-9}$. The slightly sublinear relationship between $\Sigma_{\dot{M}_\mathrm{out}}$ and $\Sigma_\mathrm{SFR}$ suggests that more massive systems with stronger $\Sigma_\mathrm{SFR}$ have more difficulty driving outflows relative to their masses and star formation activity, which is also in accordance with our analysis of the mass loading factors (Fig. \ref{fig:sfr_eta_gamma}). In the bottom panel of Fig. \ref{fig:of_sf}, we present the same analysis as above for only the gas in the warm ionised phase. The surface density of the warm ionised gas outflow shows a tighter correlation with $\Sigma_\mathrm{SFR}$ with $R_\mathrm{WIM} = 0.91$ and scales more strongly with $\alpha_\mathrm{WIM} = 1.40 \pm 0.24$. Galactic outflows driven by the hot gas phase for the highest gas surface densities can accelerate less dense ionised gas more efficiently, and the amount of ionised gas launched from the midplane ISM increases exponentially with $\Sigma_\mathrm{SFR}$. However, the colder and more mass-loading outflow, which appears to be a CR-supported outflow, is less efficiently accelerated in more massive systems. This may explain the sublinear correlation between $\Sigma_{\dot{M}_\mathrm{out}}$ and $\Sigma_\mathrm{SFR}$. In general, it seems evident that all systems deplete their star-forming gas reservoirs more through galactic outflows than through conversion into stars, as indicated by $\eta > 1$. 

\begin{table*}
	\centering
	\caption{Global star formation and outflow properties. The given values of the star formation rate surface density, $\Sigma_\mathrm{SFR}$, the mass loading factors at a height of $|z| = 1\,\mathrm{and\,} 2\,\mathrm{kpc}$, $\eta_\mathrm{1kpc}$ and $\eta_\mathrm{2kpc}$, and the energy loading factors, $\gamma_\mathrm{1kpc}$ and $\gamma_\mathrm{2kpc}$ are the medians of the total star-forming evolution (that is starting from $t_\mathrm{SFR}$ to $t_\mathrm{end}$) with \nth{25} and \nth{75} percentiles as lower and upper bounds. The net mass flow rate, $\dot{M}_\mathrm{net}$, is calculated as the total integrated mass outflow rate minus the total integrated mass inflow rate through the respective boundaries.}
	\begin{tabular}{lccccccc}
		\hline
		Name                      & $\Sigma_\mathrm{SFR}$                      & $\eta_\mathrm{1kpc}$ & $\eta_\mathrm{2kpc}$ & $\gamma_\mathrm{1kpc}$ & $\gamma_\mathrm{2kpc}$ & $\dot{M}_\mathrm{net}^\mathrm{1kpc}$ & $\dot{M}_\mathrm{net}^\mathrm{2kpc}$ \\
		                          & [$10^{-2}$ M$_\odot$ yr$^{-1}$ kpc$^{-2}$] &                      &                      &                        &                        & [$10^{-3}$ M$_\odot$ yr$^{-1}$]      & [$10^{-3}$ M$_\odot$ yr$^{-1}$]      \\
		\hline
		$\mathbf{\Sigma010}$      & $0.3^{0.5}_{0.1}$                          & $2.8^{6.7}_{0.1}$    & $0.8^{2.5}_{0.1}$    & $0.13^{0.24}_{0.01}$   & $0.09^{0.20}_{0.02}$   & 2.403                                & 1.163                                \\
		$\mathbf{\Sigma030}$      & $1.6^{1.9}_{0.2}$                          & $2.8^{4.5}_{1.0}$    & $1.3^{2.4}_{0.3}$    & $0.12^{0.22}_{0.06}$   & $0.09^{0.18}_{0.02}$   & 6.548                                & 3.573                                \\
		$\mathbf{\Sigma050}$      & $4.6^{5.0}_{1.0}$                          & $1.8^{4.5}_{0.5}$    & $1.0^{3.1}_{0.1}$    & $0.07^{0.16}_{0.02}$   & $0.06^{0.13}_{0.00}$   & 21.525                               & 13.514                               \\
		$\mathbf{\Sigma100}$      & $14.6^{17.9}_{1.0}$                        & $0.9^{2.7}_{0.4}$    & $0.9^{2.5}_{0.1}$    & $0.10^{0.35}_{0.04}$   & $0.08^{0.31}_{0.01}$   & 57.185                               & 42.252                               \\
		$\mathbf{\Sigma010^\dag}$ & $0.2^{0.2}_{0.1}$                          & $0.03^{0.6}_{0.0}$   & $0.0^{0.2}_{0.0}$    & $0.00^{0.01}_{0.00}$   & $0.00^{0.01}_{0.0}$    & 0.017                                & -0.003                               \\
		$\mathbf{\Sigma100^\dag}$ & $21.2^{37.7}_{0.4}$                        & $1.0^{6.1}_{0.0}$    & $0.23^{2.6}_{0.0}$   & $0.04^{0.15}_{0.0}$    & $0.01^{0.09}_{0.0}$    & 30.023                               & 27.256                               \\
		\hline
	\end{tabular}
	\label{tab:sfr_of_global}
\end{table*}

We summarise the global outflow properties of our ISM simulations in Table \ref{tab:sfr_of_global}. Quoted values for the star-formation rate surface density and loading factors are their characteristic values, which we define as the median of the total time evolution with active star formation. Runs that incorporate CR feedback have a 2 to 3.5 higher total outflow rate than the inflow rate, i.e. the gas flow rate through the plane in the direction toward the midplane, with a net mass flow rate, $\dot{M}_\mathrm{net} \equiv \dot{M}_\mathrm{out} - \dot{M}_\mathrm{in} \approx 1 - 60 \times 10^{-3}\,\mathrm{M_\odot\,yr^{-1}}$.

\subsection{Phase structure of the outflow}\label{sec:PhaseStructureOfTheOutflow}

\begin{figure*}
	\centering
	\includegraphics[width=.94\linewidth]{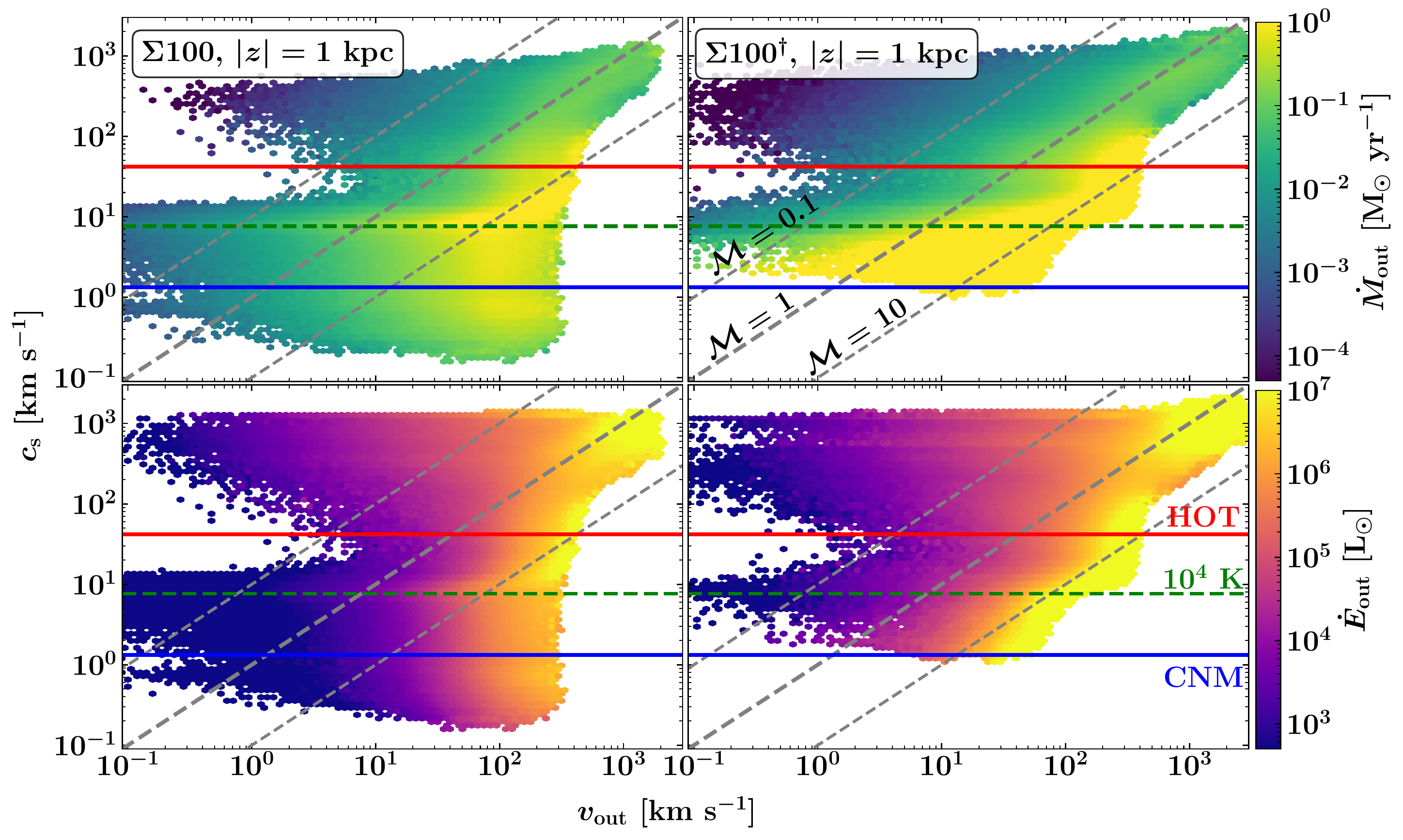}\\
	\includegraphics[width=.94\linewidth]{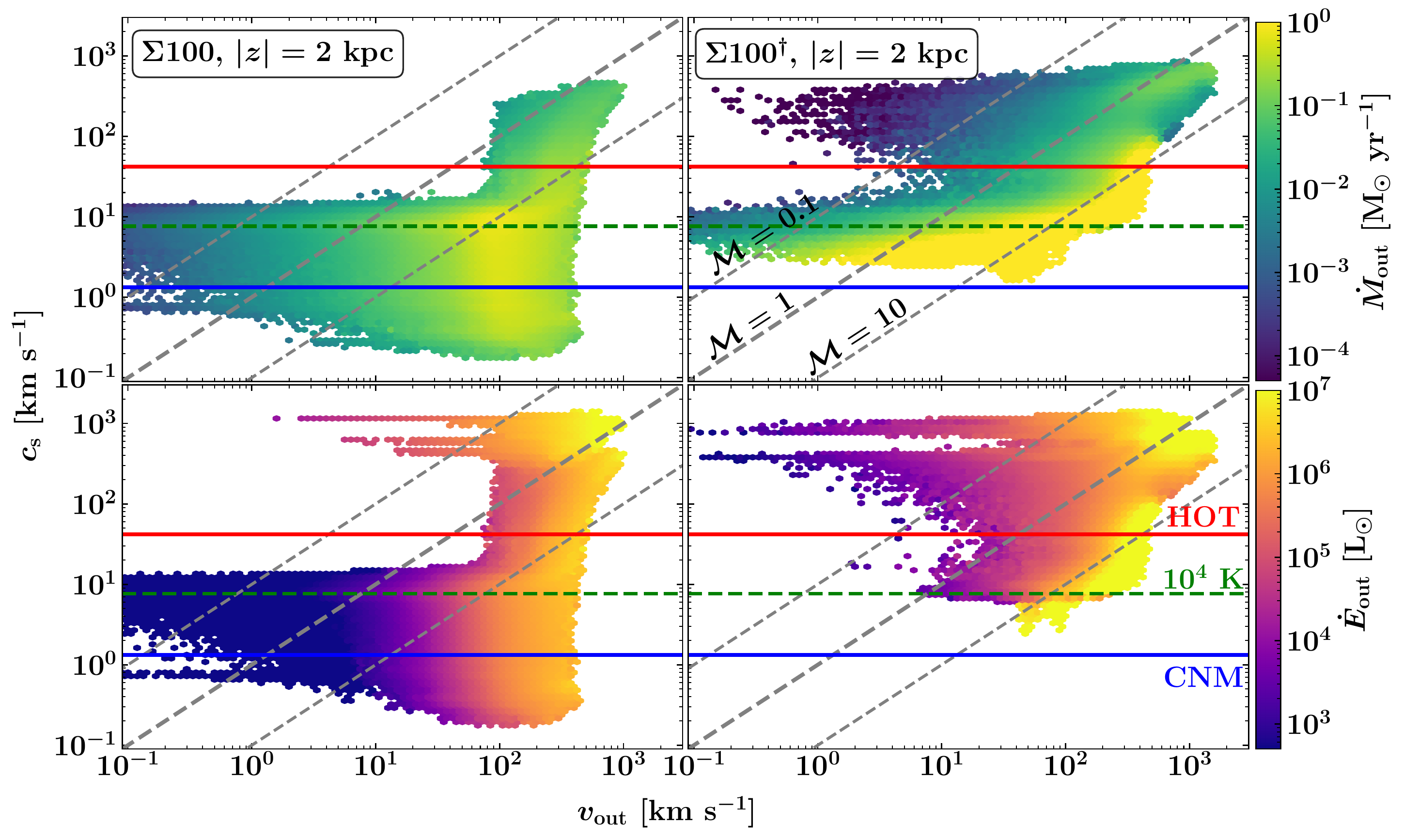}
	\caption{Sound speed $c_\mathrm{s}$ vs. outflow speed $v_\mathrm{out}$ for outflowing gas measured at $|z|$ = 1 (top) and $ z= 2\,\mathrm{kpc}$ (bottom) for the simulation with an initial gas surface density of $\Sigma_\mathrm{gas} = 10\,\mathrm{M_\odot}\,\mathrm{pc}^{-2}$, with (left-hand side) and without (right-hand side) CRs. The respective top panels are colour-coded with the mass outflow rate, $\dot{M}_\mathrm{out}$, and the bottom panels with the total energy outflow rate, $\dot{E}_\mathrm{out}$. Included in this analysis are all snapshots from the total time evolution with $\eta_ > 1$ or $\gamma > 0.1$, respectively. The horizontal coloured lines indicate transitions between gas phases with the cold neutral medium ($T < 300 \,\mathrm{K}$, blue line) and the hot gas phase ($T > 3 \times 10^5\,\mathrm{K}$, red line). The green dashed line indicates a temperature of $T = 10^4\,\mathrm{K}$, the typical temperature within HII regions. The grey diagonal lines indicate constant Mach numbers of $\mathcal{M} = 0.1, 1, 10$ from top to bottom. In simulations with CRs (left panels) mass and energy are carried in all thermal gas phases. Without CRs (right panels) the mass is carried by the warm-hot phase ($10^4 K < T < 3 \times 10^5 K$) and the energy is mostly carried by the hot phase. CR-supported outflows are multiphase in nature and we do not obtain any outflow in the cold gas phase ($T < 300\,\mathrm{K}$) without CRs.}
	\label{fig:vouts_100}
\end{figure*}

In this section, we quantify the multiphase nature of the outflow and the impact of CRs in a more detailed way.
In Fig. \ref{fig:vouts_100} and Fig. \ref{fig:vouts_010}, we show phase diagrams inspired by similar plots in \citet{Kim2020b} of the sound speed, $c_\mathrm{s}$, as a function of the velocity of the outflowing gas, $v_\mathrm{out}$, for $\Sigma100$ and $\Sigma100^\dagger$ (Fig. \ref{fig:vouts_100}), and for $\Sigma010$ and $\Sigma010^\dagger$ (Appendix Fig. \ref{fig:vouts_010}). In the four upper panels, the respective quantities are measured at $|z| = 1\,\mathrm{kpc}$, and at $|z| = 2\,\mathrm{kpc}$ in the four lower panels. The left-hand side of both figures depicts simulations that include CRs, and the right-hand side omits CR feedback. We colour-coded the data once by the mass outflow rate (upper panels in each subplot) and once by the total energy outflow rate (lower panels in each subplot). The underlying data encompass all snapshots of the complete time evolution of each simulation in which an unambiguous outflow is measurable. We define this criterion as all times with mass loading $\eta > 1$ or energy loading $\gamma > 0.1$, which encompasses between 40 and 70 per cent of the star-forming simulated time in the case of the mass loading for the simulations including CRs (10 - 36 per cent for the models without CRs) and 20 to 30 per cent of the star-forming simulated time in the case of the energy loading (5 - 20 per cent without CRs). We further indicate the different thermal gas phases with the cold neutral medium ($T < 300 \,\mathrm{K}$, blue line) and the hot gas phase ($T > 3 \times 10^5\,\mathrm{K}$, red line). The green dashed line indicates a temperature of $T = 10^4\,\mathrm{K}$, the typical temperature within the existing HII regions in our simulations. The thick grey dashed diagonal line indicates a constant Mach number of $\mathcal{M} = 1$. The thinner dashed line to the right (left) indicates a Mach number of $\mathcal{M} = 10\,(0.1)$.

The difference in the phase structure between a CR-supported outflow and an outflow purely driven by thermal stellar feedback is striking. The mass and energy outflow through both $|z| = 1\,\mathrm{kpc}$ and $|z| = 2\,\mathrm{kpc}$ encapsulates all thermal gas phases and exhibits a wide range of outflow velocities. Without the support of CRs, there exists no cold gas phase in the outflow, and we have a clear cutoff at temperatures below $T < 10^4\,\mathrm{K}$ in the outflow. 
Moving to lower surface densities under solar neighbourhood conditions, the results change slightly quantitatively but not qualitatively (see Appendix Fig. \ref{fig:vouts_010}). Again, there is no cold gas phase outflow present without CRs. Most of the mass, between 50 and 90 per cent, is carried away by the warm gas phase (see Appendix Table \ref{tab:of_phase}) while the outflow energetics are dominated by the hot gas phase (75 to 95 per cent). At the $|z| = 1\,\mathrm{kpc}$ boundary, the outflowing gas moves mainly supersonically, about $\sim 20\,\mathrm{per\,cent}$ being hypersonic (Mach number $\mathcal{M} > 10$). This ratio increases when moving further toward $|z| = 2\,\mathrm{kpc}$ up to $\sim 30\,\mathrm{per\,cent}$ of the outflowing gas being hypersonic. The same is also reflected in the characteristic mass-weighted outflow velocity of the gas, which increases between $v_\mathrm{out}^\mathrm{1kpc} \approx 30 - 65\,\mathrm{km\,s^{-1}}$ at $|z| = 1\,\mathrm{kpc}$ and $v_\mathrm{out}^\mathrm{2kpc} \approx 35 - 90\,\mathrm{km\,s^{-1}}$ at $|z| = 2\,\mathrm{kpc}$. We see a clear trend in increasing the mass-weighted outflow velocities at both boundaries with the initial gas surface density (see Table \ref{tab:of_velocity} for quantitative results). In addition to CRs, there is no additional acceleration agent at work between $|z| = 1\,\mathrm{and}\,2\,\mathrm{kpc}$. The reason for the measured increase in outflow velocity between the two heights is the fact that the slower-moving cold gas component that manages to pass through $|z| = 1\,\mathrm{kpc}$ dies off along the way to larger heights and mainly the hypersonically moving warm and hot gas component reaches $|z| = 2\,\mathrm{kpc}$. However, only the hottest outflow components reach escape velocities high enough to leave the gravitational potential of a Milky Way-like system \citep[compared to the Milky Way escape velocity of $v_\mathrm{esc} \approx 550\,\mathrm{km\,s^{-1}}$ ][]{Kafle2014}.
	
\section{Gas kinematics}\label{sec:kinematics}
\begin{figure}
	\centering
	\includegraphics[width=.99\linewidth]{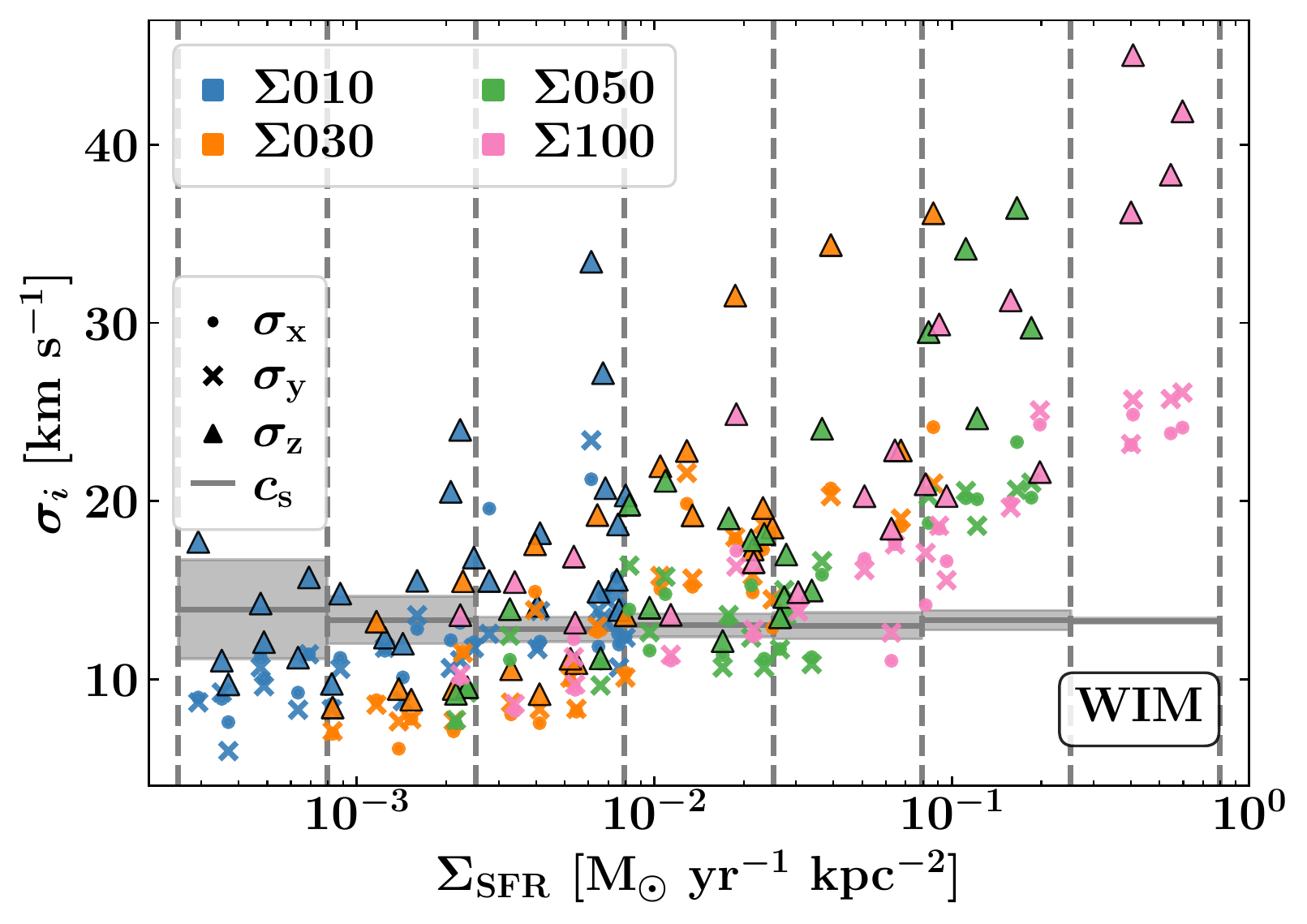}
	\caption{Mass-weighted line-of-sight velocity dispersion, $\sigma$, of the warm ionised medium (WIM) as a function of star formation rate surface density, $\Sigma_\mathrm{SFR}$ for all simulated environments including CRs. The data is sampled in $10\,\mathrm{Myr}$ bins. Different symbols indicate the velocity dispersions along the three major axes (dots: $\sigma_\mathrm{x}$, crosses: $\sigma_\mathrm{y}$, triangles: $\sigma_\mathrm{z}$). We divide the total data set into seven $\Sigma_\mathrm{SFR}$ bins of equal width in log space (grey dashed vertical lines). In each bin, the midplane-parallel velocity dispersion components $\sigma_\mathrm{x}$ and $\sigma_\mathrm{y}$ are of equal magnitude, with the perpendicular component, i.e. the outflow direction, $\sigma_\mathrm{z}$, being systematically larger (see Table Appendix \ref{tab:sigma_sfr}). We indicate the mass-weighted average sound speed, $c_\mathrm{s}$, of the WIM in each of the $\Sigma_\mathrm{SFR}$ bins with a grey horizontal line and 1$\sigma$ standard deviation as a grey-shaded area.}
	\label{fig:turbdisp_3D}
\end{figure}
	
In Fig. \ref{fig:turbdisp_3D}, we analyse the velocity dispersion in the warm ionised medium along each major axis, ($x$, $y$, $z$),  within $\Sigma_\mathrm{SFR}$ bins with a width of 0.5 dex. Data points are 10 Myr time averages. The different colours represent the four different CR simulations, and the symbols represent the velocity dispersion along the different axes ($\sigma_\mathrm{x}$: dots, $\sigma_\mathrm{y}$: crosses, $\sigma_\mathrm{z}$: triangles). The results are also tabulated in Appendix Table \ref{tab:sigma_sfr}. All dispersions increase with $\Sigma_\mathrm{SFR}$. There is a systematic difference between the velocity dispersions parallel to the midplane ($x$, $y$) and the velocity dispersion parallel to the outflow, i.e. $z-\,\mathrm{axis}$. In general, $\sigma_\mathrm{z}$ is larger by up to a factor of $\sim 2$ in each $\Sigma_\mathrm{SFR}$ bin, while $\sigma_\mathrm{x}$ and $\sigma_\mathrm{y}$ are of equal magnitude. The 1D velocity dispersions increase with increasing star formation activity. The equipartition between $\sigma_\mathrm{x}$ and $\sigma_\mathrm{y}$ is a clear indicator of turbulent motion. The increased velocity dispersion along the $z$-axis can be attributed to the outflow driven by star formation. The data suggest a clear correlation between the velocity dispersion in the ionised gas and the star formation rate surface density. However, this correlation does not necessarily imply causation. An increased gravitational collapse in more massive systems is likely to also boost star formation and an increased velocity dispersion along the vertical axis. Star formation and velocity dispersion could both correlate with gravitational instability and, therefore, seem to correlate with each other.

\begin{figure}
	\centering
	\includegraphics[width=.99\linewidth]{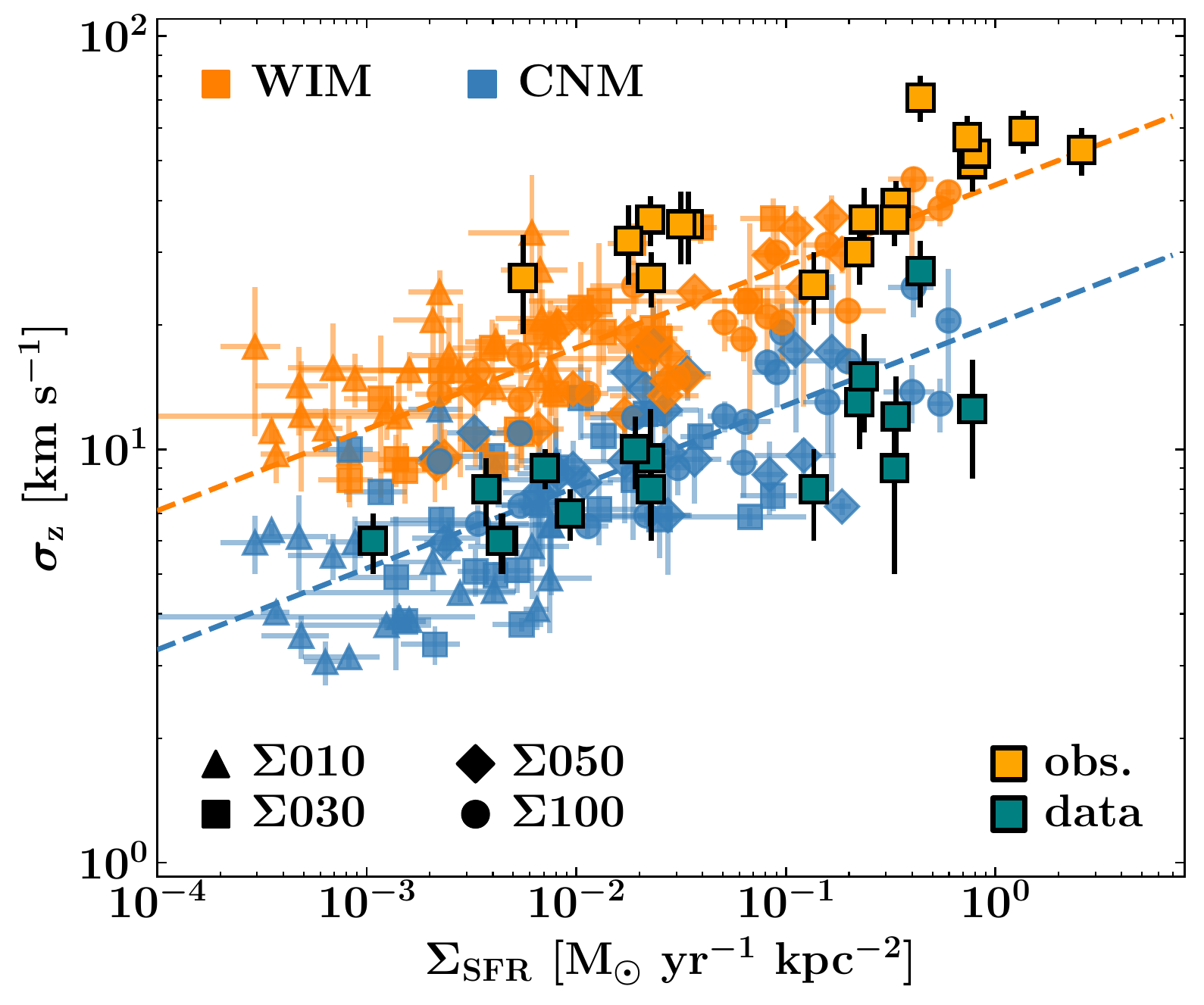}
	\caption{Mass-weighted line-of-sight velocity dispersion in $z$-direction of the warm ionised (orange) and the cold neutral medium (blue) of all simulated ISM environments including CRs as a function of the star formation rate surface density, $\Sigma_\mathrm{SFR}$. Different symbols indicate simulations with different initial conditions. The data is sampled in $10\,\mathrm{Myr}$ bins. Observational data of local star-forming galaxies, indicated with coloured squares, are taken from \citet{Leroy2013} and \citet{Girard2019, Girard2021}. The dashed lines show a power law least squares fit of the scattered data. A constant offset between the ionised and cold neutral gas phase velocity dispersion hints at the coexistence of a thin cold neutral gas disk and a thick ionised gas disk in galaxies across one order of magnitude in surface density.}
	\label{fig:sigma_sfr}
\end{figure}
	
Finally, we compile the vertical velocity dispersion, $\sigma_\mathrm{z}$, as a function of $\Sigma_\mathrm{SFR}$ for the warm ionised and cold neutral medium of all simulations that include CR transport and compare with observations (see Fig. \ref{fig:sigma_sfr}. Data from the different simulations are marked with different symbols, whereas the colour of the marker indicates the gas phase. Data points are averaged over 10 Myr bins and we also include a 1$\sigma$ standard error in $\sigma_\mathrm{z}$ and $\Sigma_\mathrm{SFR}$. The observational values from local star-forming galaxies, indicated by squares with a thick black outline, are taken from \citep{Leroy2008, Girard2019, Girard2021}. For the WIM, the observational data are derived mainly from H$\alpha$ and, for the CNM, from CO emission. The observational data agree very well with our numerical results. We also compare with additional observations of the ionised and molecular gas velocity dispersions in galaxies with high redshift \citep[$z > 2$ ][]{Tacconi2018, Molina2019, Ubler2019} and find similar agreement with our data. Nevertheless, the observed high redshift systems might be very different from our initial models in terms of galactic environment, metallicity, gravitational potential and star formation histories, etc., and therefore we do not include this data in our comparison.

Simulation data and observations suggest that there exists a constant offset between the WIM and CNM velocity dispersions, indicating the coexistence of a thin molecular gas disk and a thicker ionised gas disk, as already proposed by \citet{Girard2021}. We follow a similar analysis as in \citet{Girard2021} and fit our numerical data for both the WIM and the CNM combined in log-log space with a power law of the kind
\begin{equation}
	\mathrm{log}_{10} \left(\frac{\sigma}{\mathrm{km\,s^{-1}}}\right) = a \times \mathrm{log}_{10} \left(\frac{\Sigma_\mathrm{SFR}}{\mathrm{M_\odot\,yr^{-1}\,kpc^{-2}}}\right) + b,
\end{equation}
which yields a slope of $a = 0.20 \pm 0.02$. We then proceed and fit both WIM and CNM individually with the fixed slope $a$ to determine the offsets $b_i$. We get $b_\mathrm{WIM} = 1.64 \pm 0.01$ and $b_\mathrm{CNM} = 1.30 \pm 0.02$. These fits are indicated in Fig. \ref{fig:sigma_sfr} as dashed lines (orange for WIM and blue for CNM). Now we can calculate the constant offset between those two fits as $(0.34 \pm 0.03)\,\mathrm{dex} \equiv 2.19 \pm 1.07$. Calculating the offset directly as the mean value of the ratio of the two velocity dispersions yields $\sigma_\mathrm{WIM}\, /\ \sigma_\mathrm{CNM} \approx 2.33 \pm 0.89$ (with 1$\sigma$ standard deviation). 

\section{ISM structure}\label{sec:ISMstructure}

We conclude our analysis by quantifying the volume filling fractions (VFF) and the mass fractions (MF) of the different thermal gas phases (HOT: $T > 3\times10^5\,\mathrm{K}$; Warm Ionised Medium: $300 < T \leq 3\times10^5\,\mathrm{K}$, ionisation parameter $\chi > 0.5$; Warm Neutral Medium: $300 < T \leq 3\times10^5\,\mathrm{K}$, ionisation parameter $\chi \leq 0.5$; Cold Neutral Medium: $T \leq 300\,\mathrm{K}$).

\begin{figure}
	\centering
	\includegraphics[width=.99\linewidth]{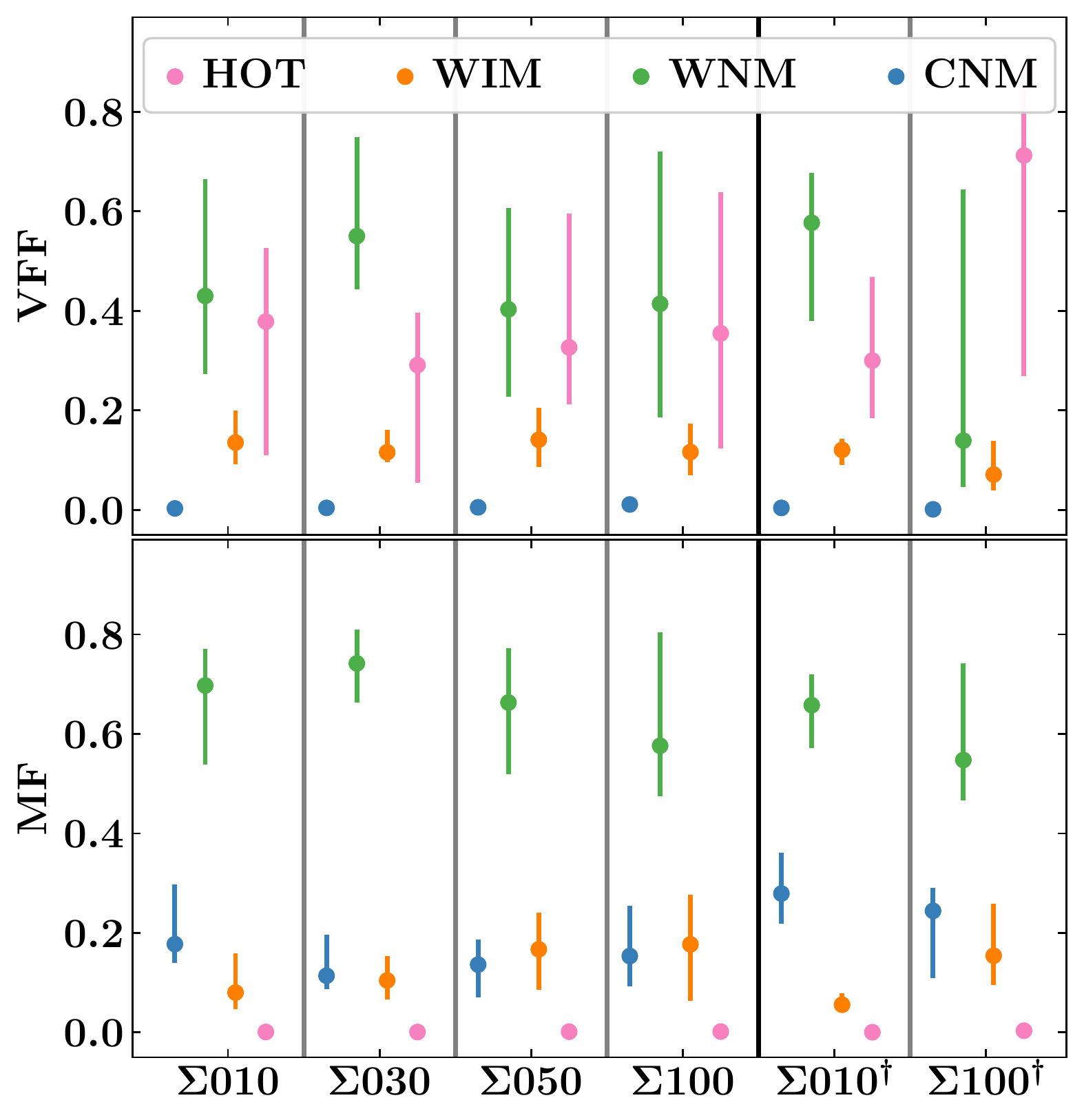}
	\caption{Characteristic values (i.e. median values from $t_\mathrm{SFR}$ until the end of the simulation) of the VFF and MF for the different gas phases. Error bars indicate the \nth{25} and \nth{75} percentile. Omitting the injection and acceleration of CRs increases the cold neutral medium mass fraction slightly. Systematic trends between the different surface densities are not present.}
	\label{fig:mf_vff}
\end{figure}

In Fig. \ref{fig:mf_vff}, we show the characteristic values (i.e. median values from $t_\mathrm{SFR}$ until the end of the simulation) of the VFF and MF for the different gas phases. Error bars indicate the \nth{25} and \nth{75} percentiles. 
On a global scale, the different gas surface density runs do not differ strongly in their VFF. The median WNM VFF ranges between $\sim 40 - 50\,\mathrm{per\,cent}$ and the HOT VFF between $\sim 30-40\,\mathrm{per\,cent}$, with the exception of $\Sigma100^\dagger$ due to the short simulated time (see also Appendix Table \ref{tab:ISM}). We draw similar conclusions for the mass fractions as in the comparison of the volume-filling fractions. There are only moderate differences between the runs with different initial surface densities with respect to the phase structure of the ISM. We find that the median MF of the CNM lies between $\sim 10-20\,\mathrm{per\,cent}$ \citep[which is not in agreement with observational estimates of $\sim 40\,\mathrm{per\,cent}$, see e.g. ][]{Tielens2005} and does not increase systematically with increasing gas surface density. A potential reason why we underestimate the mass of the cold gas phase is the presence of sink particles, which represent the star cluster in our simulations. Although we achieve realistic star formation efficiencies (see Section \ref{sec:StarFormationAndOutflows}) a huge portion of cold gas that potentially would form molecular gas gets heated up by stellar feedback. To avoid the creation of numerical artefacts, we have to inject the thermal energy of the SN feedback and the momentum feedback of the stellar winds into a region with a radius of $\sim 3$ grid cells, which equals $\sim 12\,\mathrm{pc}$ at our spatial resolution. Stellar feedback is injected into spheres with a 12 pc radius, whereas young star clusters have a physical size of $\sim 1\,\mathrm{pc}$. As a result, we inject energy and momentum into a too large volume and heat up too much dense cold gas. In reality, most of the cold gas is not destroyed but moved from the cluster vicinity to the surface of the formed bubble. A solution to this might only be to increase the numerical resolution and therefore decrease the sink particle accretion and feedback injection radius. This would result in a better resolved cold and dense gas structure. However, increasing the spatial resolution is numerically challenging and increasing the resolution to the point where molecular gas formation converges \citep[$\sim 0.1$ pc, see e.g. ][]{Seifried2020} is not feasible with stratified disk ISM simulations that try to include all major stellar feedback processes.

\section{Discussion}\label{sec:discussion}

We give a brief summary of our findings and discuss them in context.

\subsection{The characteristics of galactic outflows}

We find characteristic CR-supported mass loading factors that systematically decrease with the initial gas surface density of the system from $\eta_\mathrm{1kpc} = 2.8$ for $\Sigma_\mathrm{gas} = 10\,\mathrm{M_\odot\,pc^{-2}}$ down to $\eta_\mathrm{1kpc} = 0.9$ for $\Sigma_\mathrm{gas} = 100\,\mathrm{M_\odot\,pc^{-2}}$. The relative importance of CRs in driving galactic outflows decreases with $\Sigma_\mathrm{SFR}$. The peak values for $\eta_\mathrm{1kpc}$ are comparable between the different surface density systems (see Fig. \ref{fig:sfr_eta_gamma} and Table \ref{tab:sfr_of_global}). About $\sim 20\,\mathrm{per\,cent}$ of the mass outflow is in the hot gas phase($T > 3\times10^5\,\mathrm{K}$), another $\sim 20\,\mathrm{per\,cent}$ in the cold neutral medium ($T < 300\,\mathrm{K}$) and the remaining $\sim 60\,\mathrm{per\,cent}$ in the warm neutral and warm ionised gas phase (see Table \ref{tab:of_phase}). The omission of anisotropic CR diffusion in our models drastically changes this picture and leads to a lack of a cold and potentially molecular gas in the outflow altogether. Most of the energy is transported away from the midplane ISM in the hot gas phase ($\sim 80-90\,\mathrm{per\,cent}$) with energy loading factors of $\gamma_\mathrm{1kpc} \approx 0.1$ for all surface densities when CRs are included. There are no systematic differences detectable in the thermal composition of the outflowing gas between the models that include CRs. The gas leaving the ISM in the midplane through $|z| = 1\,\mathrm{kpc}$ travels mostly with supersonic velocities ($\sim 70\,\mathrm{per\,cent}$) or even hypersonic with Mach numbers $\mathcal{M} > 10$ ($\sim 20\,\mathrm{per\,cent}$). The relative amount of gas moving with hypersonic velocities increases even further with $\sim 30\,\mathrm{per\,cent}$ at $|z| = 2\,\mathrm{kpc}$. However, characteristic mass-weighted outflow velocities range between $\sim 30\,\mathrm{km\,s^{-1}}$ in $\Sigma010$ and up to $\sim 90\,\mathrm{km\,s^{-1}}$ in $\Sigma100$, and most of the gas is likely to fall back onto the midplane ISM and establish a long-lived fountain flow system (see Appendix Table \ref{tab:of_velocity}). These results clearly show the multiphase nature of galactic outflows and underline the importance of including CRs when modelling galactic outflows.

Our numerical predictions on the strength of the ionised gas outflow are consistent with the work of \citet{ForsterSchreiber2019}, who use the KMOS$^\mathrm{3D}$ survey to analyse the demographics and properties of the ionised gas outflows in a redshift range of $z = 0.6 - 2.7$. This sample stretches over a wide range of quiescent and starburst systems with and without active galactic nuclei (AGN). However, in their analysis, they can distinguish star formation-driven and AGN-driven galactic outflows and quote an ionised gas mass loading of $\eta_\mathrm{ion} \approx 0.1 - 0.2$. We find time-averaged warm-ionised gas mass loadings at $|z| = 1\,\mathrm{kpc} = (0.80, 0.17, 0.16, 0.12)$ for ($\Sigma010$, $\Sigma030$, $\Sigma050$, $\Sigma100$), respectively. \citet{ForsterSchreiber2019} argue that those low mass loading factors would be insufficient to regulate star formation via galactic outflows and argue that they can only be a lower limit for the total mass, momentum, and energy outflow rates, which we can confirm in our simulations. This highlights the necessity of simulations of galactic outflow systems that incorporate all relevant driving agents for a fully multiphase galactic outflow to match against observations.

Using the DUVET survey, \citet{Chu2022} study the ionised and molecular outflows of the starburst galaxy IRAS08339+6517 at $z \sim 0.02$, which shows a range of star formation surface densities between $\Sigma_\mathrm{SFR} \approx 0.01 - 10\,\mathrm{M_\odot\,yr^{-1}\,kpc^{-2}}$ across resolution elements of a few hundred parsecs size. This system is to some extent comparable to the $\Sigma100$ simulation presented in this work. They estimate a total mass loading of $\eta\sim2-10$ but caution about uncertainties in estimating total mass outflow rates from ionised gas outflow rates. We find a median mass loading of $\eta_\mathrm{1kpc} = 0.9$ for $\Sigma100$. Furthermore, they find a slight anticorrelation between $\Sigma_\mathrm{SFR}$ and $\eta$, as well as a correlation between $\Sigma_\mathrm{SFR}$ and $v_\mathrm{out}$, similar to our findings. In a follow-up study of the same system \citep{Chu2022a}, they determine a relation of $\Sigma_{\dot{M}_\mathrm{WIM}} \propto \Sigma_\mathrm{SFR}^{1.06\pm0.10}$ under the assumption that the ionised outflow mass scales with the total outflow mass. We find a steeper relation between the surface density of the ionised gas outflow ($\alpha = 1.40 \pm 0.24$) but a slightly sublinear relation when we take into account the total gas mass of the outflow ($\alpha = 0.81 \pm 0.12$, see Fig. \ref{fig:of_sf}). It becomes clear that the assumption that the total gas outflow scales with the ionised gas outflow can lead to false estimates and underlines the importance of resolved simulations of galactic outflows to gauge observational estimates.

The importance of the multiphase nature of galactic outflows is also found in the numerical work by \citet{Kim2020b} who analyse the galactic outflows launched in the \textsc{\textsc{tigress}} \citep{Kim2017} simulation suite, which very closely resembles our \textsc{silcc} setup. They find that most of the mass is launched as galactic outflows in the "cool" phase, which they define as $T < 2\times10^4\,\mathrm{K}$. This aligns with our definition of the warm medium. Additionally, the "hot" ($T > 5\times10^5\,\mathrm{K}$, very close to our definition of the hot gas phase), carries most of the outflowing energy. Importantly, the galactic outflows in their setup are solely launched by SNe since they do not model CR transport. This results in the cold gas phase ($T < 300\,\mathrm{K}$) missing in the outflow. The outflow velocity distribution detected in their "hot" and "cool" phases is broad, especially in the $T ~ 10^4\,\mathrm{K}$ gas, which aligns with our findings (see Fig. \ref{fig:vouts_100}). This broad outflow velocity distribution leads to the fact that within a specific thermal gas phase, the amount of gas which can escape the galactic potential differs and cannot be accurately modelled just by the thermal condition of the gas. This becomes even more evident when a CR-supported outflow is considered, in which gas can be lifted at very moderate velocities compared to the ballistic outflows driven by SNe.

\subsection{The role of CRs}\label{sec:crmodel}
 
We see a strong influence of CRs on the shape and characteristics of galactic outflows. Without CRs, we cannot transport cold (and possibly molecular) gas out of the midplane ISM. Furthermore, extended cold and warm outflows move with a wide range of velocities, ranging from subsonic ($\mathcal{M} < 1$) to hypersonic ($\mathcal{M} > 10$). An outflow supported by CRs is multiphase in nature, even up to heights above $|z| = 2\,\mathrm{kpc}$, while outflows driven without an additional CR pressure gradient are two-phase (warm and hot) at $|z| = 1\,\mathrm{kpc}$ and only single-phase (hot) at $|z| = 2\,\mathrm{kpc}$. However, the efficiency with which CRs can drive and sustain outflows is dependent on the surface density of the system. In solar neighbourhood conditions, we see a striking increase in mass loading when CRs are included compared to non-CR models (Fig. \ref{fig:sfr_eta_gamma}: right panel and Fig. \ref{fig:vouts_100}). In more massive environments (such as $\Sigma100$), the impact of CRs is less pronounced due to higher hadronic energy losses, and the main driver of strong outflows remains the volume-filling hot gas created by overlapping SNe. 

\citet{Simpson2016} study the impact of CR pressure on accelerating galactic outflows with idealised stratified disk simulations and report a strong increase in total mass outflow with a smoother density structure in accordance with our results \citep[see also][]{Girichidis2018}. Similar conclusions are reached by \citet{Dashyan2020} who simulate isolated dwarf galaxy systems with the AMR code \textsc{ramses} with different models for CR diffusion with varying diffusion coefficients as well as CR streaming. They also see a clear increase in galactic outflow generation when anisotropic CR diffusion (similar to what is in our models) is at play. Furthermore, \citet{Chan2022} have investigated the impact of CR in the FIRE-2 cosmological simulation suite and find that CRs increase the amount of warm ($T\sim10^4\,\mathrm{K}$) gas in the outflow and their disk-halo interface becomes dominated by the volume filling warm-hot gas phase ($T\sim 2\times10^4 - 5\times10^5\,\mathrm{K}$). Without CRs, most of their outflow originates from hot superbubbles, similar to our results. CRs are likely to mediate the evolution of the midplane ISM, depending on the gas properties at their injection site \citep{Simpson2023}. The presence of CRs sets a minimum pressure floor, and CRs prevent a thermal runaway ISM with a spatially random injection of SNe at a fixed SN injection rate \citep[see also][and references therein for the effect of different placements of SNe on the ISM at a fixed injection rate]{Naab2017}.

Our models so far do not include CR streaming and we do not see a strong impact of CRs on the properties of the midplane ISM. \citet{Dashyan2020} conclude that CR streaming has only a minor impact on star formation and galactic outflows, however, \citet{Wiener2017} demonstrate that CR streaming can drain a significant amount of energy from CRs, which would also reduce the ability to build up the long-lasting and far-reaching CR pressure gradient, which is responsible for the very efficient driving of galactic outflows. Toward these ends, \citet{Thomas2019} study the coupling of CR streaming with self-excited Alfv\'en waves in a self-consistent CR-MHD formulation and emphasise the importance of CR streaming to account for proper CR scattering and arrive at a realistic CR momentum density. Moreover, our CR implementation does not resolve the full CR spectrum but considers only a single GeV energy bin with an assumed steady-state spectrum. \citet{Girichidis2022} use the MHD code \textsc{arepo} to perform isolated galaxy simulations with spectrally resolved CRs, which allows for more precise modelling of CR cooling and enables energy-dependent spatial diffusion. They find that the high-energy CRs diffuse faster through the medium, which allows CR-supported outflows to be launched farther away from the galactic centre where most of the star formation and CR injection takes place. At the same time, the low-energy part of the CR spectrum leads to a smaller diffusion coefficient, which leads to saturation of the CR pressure close to star-forming regions. Due to the greatly increased numerical cost of including a full CR spectrum, which ranges from a couple of MeV to a couple of TeV in our here presented \textsc{silcc} simulations, we are limited to the so-called "grey" CR approach with a steady-state spectrum assumption. In our steady-state model, we assume a fixed diffusion coefficient along the magnetic field lines of $K_\parallel = 10^{28}\,\mathrm{cm^2\,s^{-1}}$ and of $K_\bot = 10^{26}\,\mathrm{cm^2\,s^{-1}}$ perpendicular to the magnetic field, based on observational estimates by \citet{Strong2007} and \citet{Nava2013}. 
Increasing the diffusion coefficient would most likely decrease the CR pressure gradient in the disc, which could reduce the impact of CRs on the outflows as found in \citet{Girichidis2018}. Most of the CR energy is in the form of CR protons with a particle energy of a few GeV. Thus, changing the diffusion coefficient for the bulk of the CR energy will also affect their impact. 
On one hand, a larger diffusion coefficient leads to a stronger CR flux, which therefore reduces CR energy over-densities in the midplane and therefore reduces the CR pressure gradient. A lower diffusion coefficient is therefore expected to establish a larger CR pressure gradient which possibly accelerates more outflows. However, CRs with a lower diffusion coefficient reside longer in high-density gas in which they cool more efficiently, leading to a decrease in CR pressure gradient. Nonetheless, CRs with a large diffusion coefficient can easier reach heights with more diffuse and diluted gas, which in turn makes it easier for them to accelerate the gas compared to the dense gas in the midplane ISM. Which of the above mechanisms dominates is unclear up to now. \citet{Girichidis2018} find that more outflow is accelerated by a larger pressure gradient generated through a lower diffusion coefficient but they do not see a potent effect of CR cooling. It is expected that spectrally resolved CRs as well as CR streaming might enhance the CR cooling efficiency. A similar trend can be expected by varying diffusion coefficients of a spectrally resolved CR model. Overall, the integrated differences in total outflow rate and star formation rate between a "grey" and spectrally resolved approach appear to be subtle and might not impact the overall evolution of the ISM to a great extent \citep{Girichidis2022, Girichidis2023}, whereas the difference between an ISM model with and without CR diffusion is striking.

\subsection{Origin of the velocity dispersion}

We have demonstrated a clear correlation between the velocity dispersion in the WIM/CNM and the star formation rate surface density, especially true for the velocity dispersion along the outflowing $z-$axis. There exists a systematic increase by a factor of $\sim 2$ of velocity dispersion along the outflowing axis, $\sigma_\mathrm{z}$, compared to velocity dispersions in parallel directions to the midplane, $\sigma_\mathrm{x}$ and $\sigma_\mathrm{y}$, which are of equal magnitude (Fig. \ref{fig:turbdisp_3D}). Both $\sigma_\mathrm{z}^\mathrm{WIM}$ and $\sigma_\mathrm{x,y}^\mathrm{WIM}$ scale with $\Sigma_\mathrm{SFR}$ while the dependence is slightly weaker for $\sigma_\mathrm{x,y}^\mathrm{WIM}$. When fitting a power law to the vertical velocity dispersions of the WIM and CNM as a function of $\Sigma_\mathrm{SFR}$, we find a similar slope of $a = 0.20 \pm 0.02$ and a constant offset of the vertical velocity dispersion in the warm ionised medium compared to the cold neutral medium with a factor of $2.19\pm1.07$. The plane-parallel velocity dispersions of the CNM, $\sigma_\mathrm{x,y}^\mathrm{CNM}$ have a shallower scaling with $\sigma_\mathrm{x,y}^\mathrm{CNM} \approx \Sigma_\mathrm{SFR}^{0.16\pm0.01}$. 
For star formation surface densities above $\Sigma_\mathrm{SFR} \gtrsim 1.58 \times 10^{-2}\,\mathrm{M}_\odot\,\mathrm{yr}^{-1}\,\mathrm{kpc}^{-2}$, the WIM becomes fully supersonic (see Fig. \ref{fig:turbdisp_3D}). A similar constant offset between the ionised gas velocity dispersion, corrected for thermal broadening and measured in H$\alpha$ and/or H$\beta$, and the molecular gas velocity dispersion (measured mainly in CO transitions) in local $z = 0.5 - 2.5$ galaxies compiled from literature has been found by \citet{Girard2021}. Among other results, they analyse the velocity dispersion as a function of the gas fraction and arrive at a constant offset between the ionised and molecular gas of $2.45\pm0.38$. This constant offset may hint at the coexistence of a thin molecular gas disk embedded in a thicker ionised gas disk.

The origin of the velocity dispersion is ambiguous. The correlation with $\Sigma_\mathrm{SFR}$ and the systematic increase of $\sigma_\mathrm{z}$ compared to $\sigma_\mathrm{x}$ \& $\sigma_\mathrm{y}$ suggest that the outflows driven by star formation could be the source of velocity dispersion. 
On the other hand, \citet{Krumholz2016} and \citet{Krumholz2018} have developed a theoretical model of gas in vertical hydrostatic equilibrium, which predicts that for feedback-driven turbulence in galactic discs the star formation rate would scale sharply with velocity dispersion ($\dot{M_\star} \propto \sigma_\mathrm{3D}^2$) while star formation triggered by turbulence through gravitational instabilities exhibits a shallower slope, which is in apparent agreement with observations. In this model, energy is always balanced between the energy input from stellar feedback and turbulent decay, as well as radial mass transport through the galactic disk to release gravitational energy. \citet{Ubler2019} study the velocity dispersions of ionised and atomic + molecular gas in 175 star-forming disk galaxies in the redshift range $z \sim 0.6 - 2.6$ from the KMOS$^\mathrm{3D}$ survey and find that the ionised gas velocity dispersions are $\sim 15\,\mathrm{km\,s^{-1}}$ higher on average than the atomic+molecular gas velocity dispersion, which follows our results (see Fig. \ref{fig:sigma_sfr}). Furthermore, they argue that the observed disks are only marginally Toomre-stable, which suggests that the turbulence is fuelled by gravitational instabilities, while the turbulence of stellar feedback is insufficient to explain the observed high velocity dispersions of the ionised gas. \citet{Ubler2019} find that $\geq 60\,\mathrm{per\,cent}$ of their observed galaxies show agreement with the predictions of the gravity-driven turbulence model of \citet{Krumholz2018}. 

\citet{Ejdetjarn2022} simulate isolated disc galaxies to quantify the origin of the gas velocity dispersion and study the impact observational effects such as beam smearing can have in estimating the $\sigma-\dot{M_\star}$ relation. In their experiments, they can turn off stellar feedback altogether and could not find an impact on the total gas velocity dispersion and therefore argue that galaxies self-regulate their turbulence by gravitational instabilities. However, stellar feedback significantly increases the dispersion of the ionised gas velocity by up to $\sigma_\mathrm{WIM} \sim 100\,\mathrm{km\,s^{-1}}$ in their models. They caution that beam-smearing effects can increase the observed velocity dispersion by factors of several. Furthermore, the general gas kinematics are traced differently by different gas tracers, and relying solely on the warm ionised phase (which is traced primarily by H$\alpha$) can lead to an overestimation of the total turbulent energy in the gas.
\citet{Jimenez2022} analyse $\sigma_\mathrm{z}$ in cosmological \textsc{eagle} simulations and try to isolate the origin of the velocity dispersion by turning stellar and/or AGN feedback on and off. Unfortunately, those simulations are limited by temporal and spatial resolution, which made a clear distinction between the important agents infeasible. They found the strongest correlation of $\sigma_\mathrm{z}$ at a fixed halo mass with the gas accretion rate, which at first glance would support the idea of gravitational instabilities being the most important source for the high velocity dispersion. However, \citet{Jimenez2022} argue that the highly nonlinear interaction of multiple physical processes determines the strength of the vertical turbulence and that the relative importance varies for different halo masses and redshifts. It is important to note that the numerical experiments discussed above do not include CRs in their stellar feedback models. As seen in this work, CRs are a crucial component and alter the structure of the gas velocity dispersion by their ability to drive strong outflows in low $\Sigma_\mathrm{SFR}$ regimes. Our determined slope for the vertical velocity dispersion in the WIM as a function of star formation rate surface density, $\sigma_\mathrm{z}^\mathrm{WIM} \propto \Sigma_\mathrm{SFR}^{0.20\pm0.02}$ is rather in favour of a stellar feedback-driven turbulence according to the analytic unified model of galactic disk turbulence \citep{Krumholz2018} which arrives at $\Sigma_\mathrm{SFR} \propto \sigma_\mathrm{3D}^2 \rightarrow \sigma_\mathrm{3D} \propto \Sigma_\mathrm{SFR}^{0.5}$ than of the stellar feedback plus radial transport prediction.

To further quantify the impact of $\Sigma_\mathrm{SFR}$ on the turbulent velocity dispersion, we have simulated a test model of the high gas surface density system ($\Sigma_\mathrm{gas} = 100\,\mathrm{M_\odot\,pc^{-2}}$) without any stellar feedback processes turned on, $\Sigma100$-noFB (see Appendix Fig. \ref{fig:noFB}). We find that the strong correlation between $\sigma^\mathrm{WIM}-\Sigma_\mathrm{SFR}$ vanishes (Spearman rank correlation coefficients of $R = (0.25, 0.19, 0.27)$ for $\sigma_\mathrm{x, y, z}$ with high $p$-values of (0.20, 0.34, 0.17), respectively, see Appendix Table \ref{tab:noFB}). The average vertical velocity dispersion at high star formation rate surface densities without stellar feedback is $\sim 10\,\mathrm{km\,s^{-1}}$ lower than in the case with feedback. However, $\Sigma100$-noFB generates $\sigma_\mathrm{z}^\mathrm{WIM}$ peak values above $60\,\mathrm{km\,s^{-1}}$, which is $\sim 15\,\mathrm{km\,s^{-1}}$ faster than in the feedback models. Turbulence fuelled by gravitational collapse alone can reach high velocity dispersions in the absence of stellar feedback. However, this is not a necessity and there are periods in which the WIM even enters the subsonic regime again while the star formation activity is still at a maximum (Appendix Fig. \ref{fig:noFB}). With the effects of gravitational instabilities plus stellar feedback, it is possible to consistently drive turbulence with high velocity dispersions in the WIM, which unambiguously scales with $\Sigma_\mathrm{SFR}$ and becomes fully supersonic at higher $\Sigma_\mathrm{SFR}$. However, ultimately we cannot conclusively test the impact of gravitational instabilities and radial mass transport, since we do not simulate a full rotating galactic disc, but an isolated $500 \times 500\,\mathrm{pc^2}$ wide patch with periodic boundary conditions.

\subsection{Caveats and future improvements}\label{sec:caveats}

We varied the initial gas surface density, $\Sigma_\mathrm{gas}$, of our simulated systems to model the effect of different galactic environments with increased star formation activities. Although we have changed $\Sigma_\mathrm{gas}$ and the strength of the magnetic field and ISRF accordingly, we have kept the external gravitational potential, which also includes the old stellar component and the metallicity fixed throughout all models. We have chosen to do so to maintain better control over the results as we do not have fiducial models of \textsc{silcc} simulations with those parameters varying at this point\footnote{A study about the impact of different metallicities is currently underdone by Brugaletta et al. (in prep.).}. One can easily assume that decreasing the metallicity would reduce star formation (owing to less cooling), while increasing the gravitational potential would increase star formation (due to greater gravitational collapse). The question of whether the increased star formation would drive a stronger galactic outflow or whether the additional gravitational acceleration would reduce the outflow is ambiguous. 

Massive OB runaway stars have been thought of as a possibly important component in ISM studies since they might be able to distribute stellar feedback over a larger area and into a regime in which star clusters might not penetrate. However, this component is lacking in the current work, and its impact will be studied in a future iteration. However, preliminary studies so far have shown that runaway stars have only a minuscule impact on the global evolution of star formation and galactic outflows (Rathjen et al., in preparation). More extended studies of runaway stars under simulated ISM conditions have come to the same conclusion \citep[see e.g.][]{Kim2018, Steinwandel2022}. Other studies of full disk simulations, however, have found runaway stars to give a strong boost in galactic outflows \citep[see e.g.][]{Andersson2020}. None of those studies has combined the effects of runaway stars with the transport of CRs, which we have found to be one of the main drivers of galactic outflows in lower gas surface density systems. The question of whether runaway stars can make a significant difference in the mass- and energy-loading factors of a system that includes CRs remains open for analysis.
 
We resolve the cold, dense gas phase, but we do not resolve the molecular gas, strictly speaking. The base resolution of our computational domain is $dx \approx 4\,\mathrm{pc}$. Earlier work using the \textsc{silcc} framework for ISM zoom-in simulations by \citet{Seifried2017, Seifried2020} has demonstrated that spatial resolutions down to sub-parsec scales are necessary to resolve molecular gas formation. These resolution requirements cannot be met with our setup in the foreseeable future. We use the cold gas phase ($T < 300\,\mathrm{K}$) as a proxy to compare with observations of the molecular gas.

Our models are limited to the size of our periodic domain of $500 \times 500\,\mathrm{pc^2}$ in the midplane and do not include a galactic context. Without galactic rotation and the resulting large-scale shearing motions, we are missing a crucial contribution to amplifying the interstellar magnetic field through a small-scale dynamo \citep[see][and references therein]{Beck2019}. Furthermore, we cannot model radial mass transfer, which, according to the widely accepted theoretical unified model of galactic disk turbulence \citep{Krumholz2018}, is a major source of ISM turbulence. We can make precise predictions about the highly nonlinear interactions of diverse stellar feedback processes and the nonthermal ISM and the capabilities to drive galactic outflows under various circumstances, but yet need to transfer insights and models into larger-scale contexts like full-scale isolated galactic disk and dwarf galaxy simulations at high ($dx \leq 4\,\mathrm{pc}$) spatial resolutions.

\section{Conclusion}\label{sec:conclusion}
We have presented a suite of seven stratified disk MHD simulations of a galactic patch using the \textsc{silcc} simulation framework. We vary the initial gas surface density $\Sigma_\mathrm{gas}$ of our models between $10 - 100\,\mathrm{M_\odot\,pc^{-2}}$ and achieve a wide range of star formation rate surface densities of $\Sigma_\mathrm{SFR} \approx 3\times10^{-4} - 1\,\mathrm{M_\odot\,yr^{-1}\,kpc^{-2}}$. We include early stellar feedback in the form of momentum input by stellar winds and ionising radiation with an on-the-spot radiative transfer, as well as energy input by SNe. Additionally, we include the acceleration of CRs in the remnants of SNe and model their transport with anisotropic diffusion. We follow the evolution of hydrogen and carbon chemistry with a non-equilibrium chemical network, which allows us to go down to gas temperatures of $T \approx 5\,\mathrm{K}$. 
In this study, we have focused especially on the characteristics of the galactic outflows and the kinematic signature of the midplane and the outflowing gas. The takeaway points of this work are as follows:
\begin{itemize}
	\item Galactic outflows are multiphase with a broad distribution in $v_\mathrm{out}$ even within distinct thermal phases. Most of the mass is transported away from the midplane ISM in the warm gas phase. Overall, we achieve characteristic mass loading factors between $\eta \sim 1 - 3$, with mass loading decreasing with increasing gas surface density. The peak mass loading factors are independent of the mass of the system at $\eta_\mathrm{peak} \sim 10$. Approximately $10\,\mathrm{per\,cent}$ of the SN energy injected leaves the system, while most ($\sim 80-90\,\mathrm{per\,cent}$) of this energy is transported away from the midplane ISM in the hot gas phase.
	\item The relative strength of the outflow is anti-correlated with $\Sigma_\mathrm{SFR}$ and therefore with $\Sigma_\mathrm{gas}$. The data suggest that galactic outflows are the main regulator of star formation by depleting the star-forming gas reservoir for high $\Sigma_\mathrm{SFR}$ systems, whereas at lower gas surface density systems (with lower $\Sigma_\mathrm{SFR}$) star formation is regulated by outflow only to some extent, but mainly by the direct impact of stellar feedback on star cluster scales \citep[see for a similar conclusion also][]{Rathjen2021}. In general, all systems deplete their star-forming gas reservoir rather through galactic outflows than by conversion into stars, as seen in the general trend of $\eta > 1$.
	\item Only with CRs, we obtain a realistic three-phase galactic outflow. Without CRs, galactic outflows consist of a two-phased warm-hot medium at heights of $|z| = 1\,\mathrm{kpc}$ and reduce further in thermal complexity towards $|z| = 2\,\mathrm{kpc}$, where they only consist of the hot $(T > 3\times10^5\,\mathrm{K})$ gas phase. CR-supported outflows exhibit a three-phase medium throughout heights above $|z| = 2\,\mathrm{kpc}$ with a spread in the outflow velocities of the cold gas of a couple of hundreds $\mathrm{km\,s^{-1}}$.
        \item CRs are an important agent in driving and supporting galactic outflows during long-term evolutions, even throughout periods of low star-formation activity. The relative importance of CRs affecting the magnitude of a galactic outflow compared to outflows launched from the hot gas phase created in overlapping SN superbubbles decreases as the gas surface density increases.
	\item From an observer's viewpoint, concluding mass outflow rates by only measuring ionised gas outflows can lead to an overestimation of the latter. Consequently, it is of uttermost importance to include CRs in numerical predictions of outflow properties to completely capture the full multiphase nature of galactic outflows to inform observational calibrations, as well as numerical subgrid models for cosmological simulations.
	\item The dispersions of ionised gas velocity measured over all surface densities and star formation rates correlate with $\Sigma_\mathrm{SFR}$. Furthermore, the ionised gas velocity dispersion along the outflowing axis of our computational domain, $\sigma_\mathrm{z}$, is systematically larger by a factor of $\sim 2$ compared to the dispersions of the velocity along the periodic axes parallel to the midplane, $\sigma_\mathrm{x}$ and $\sigma_\mathrm{y}$, which are otherwise of equal magnitude. The motions in the WIM transition into a regime with supersonic Mach numbers for star formation rate surface densities of $\Sigma_\mathrm{SFR} \gtrsim 1.58 \times 10^{-2}\,\mathrm{M}_\odot\,\mathrm{yr}^{-1}\,\mathrm{kpc}^{-2}$.
 
 Together, these results indicate that $\sigma_\mathrm{z}$ could be originating from outflows which are in turn driven by stellar feedback. However, this result is not unambiguous since turbulence generated by gravitational instabilities, collapse, and radial matter transport throughout the galactic disk could result in similarly strong velocity dispersions. Furthermore, gravitational instabilities also lead to more star formation, which would explain why a velocity dispersion driven by turbulence fed by gravitational instabilities also correlates with star formation activity. However, turbulence is likely to result in isotropic velocity dispersions and may not be able to explain the systematic difference between $\sigma_\mathrm{z}$ and $\sigma_\mathrm{x}\,\&\,\sigma_\mathrm{y}$. The ultimate source of the turbulent velocity dispersion in the ISM at moderate $\Sigma_\mathrm{SFR}$ is ambiguous and can be attributed to the complex interplay of stellar feedback, stellar feedback-driven and CR-supported outflows, and gravitational instabilities.
\end{itemize}

\section*{Acknowledgements}

The authors thank Deanne B. Fisher for the discussions and for providing observational data. The authors acknowledge the Gauss Centre for Supercomputing e.V. (www.gauss-centre.eu) for the pn34ma grant at SuperMUC-NG hosted by the Leibniz Supercomputing Centre (www.lrz.de). TER and DS acknowledge support by the project ''NRW-Cluster for data intensive radio astronomy: Big Bang to Big Data (B3D)'' funded through the programme ''Profilbildung 2020'', an initiative of the Ministry of Culture and Science of the State of North Rhine-Westphalia. The sole responsibility for the content of this publication lies with the authors. TN acknowledges the support of the Deutsche Forschungsgemeinschaft (DFG, German Research Foundation) under Germany’s Excellence Strategy - EXC-2094 - 390783311 of the DFG Cluster of Excellence ''ORIGINS''. SW gratefully acknowledges the European Research Council under the European Community's Framework Programme FP8 via the ERC Starting Grant RADFEEDBACK (project number 679852). TER, SW and DS further thank the Deutsche Forschungsgemeinschaft (DFG) for funding through SFB~956 ''The conditions and impact of star formation'' (SW: subproject C5 and DS: subproject C6), and SW thanks the Bonn-Cologne Graduate School. PG acknowledges funding from the European Research Council under ERC-CoG grant CRAGSMAN-646955 and from the ERC Synergy Grant ECOGAL (grant 855130). RW acknowledges the support by project 20-19854S of the Czech Science Foundation and by the institutional project RVO:67985815. The software used in this work was in part developed by the DOE NNSA-ASC OASCR Flash Centre at the University of Chicago \citep{Fryxell2000, Dubey2009}. Visualisations of the simulation results were made in part using the \textsc{yt} library for Python \citep{Turk2011}.

\section*{Data Availability}

The derived data underlying this article will be shared on reasonable request to the corresponding author. The simulation data will be made available on the \textsc{silcc} data web page: \url{http://silcc.mpa-garching.mpg.de/}.

\bibliographystyle{mnras}
\bibliography{SILCC7}

\appendix\label{sec:appendix}
	
\section{Calculation of the velocity dispersion}
 
\begin{figure}
	\centering
	\includegraphics[width=.99\linewidth]{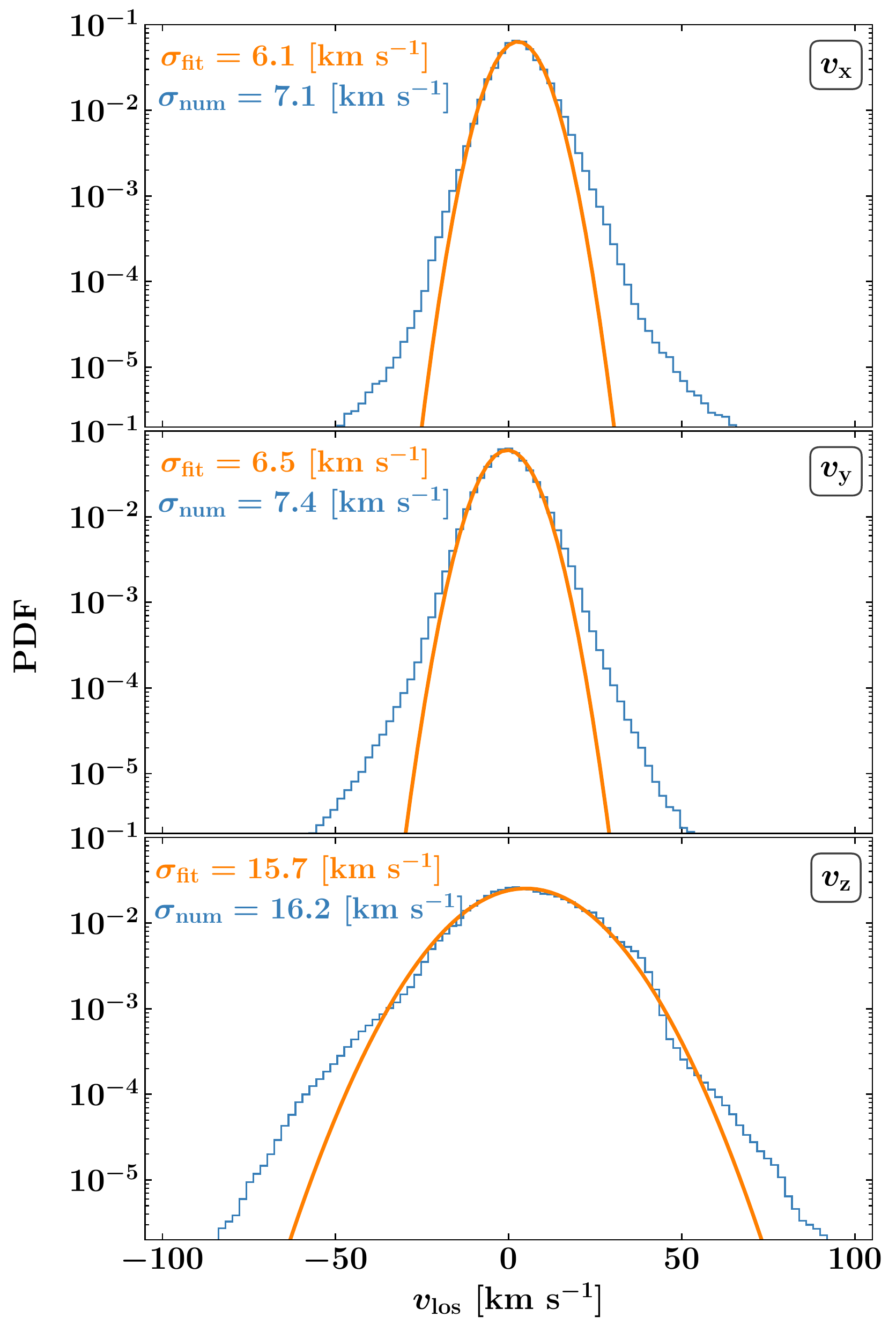}
	\caption{Velocity profile from the simulation with numerically determined velocity dispersion (blue) compared to Gaussian fit of the velocity profile with fitted dispersion (orange). Both methods agree within $\sim 10\,\mathrm{per\,cent}$ for the velocities along $x$ and $y$, and even down to $\sim 3\,\mathrm{per\,cent}$ difference for the velocity dispersion along the outflowing $z$-axis.}
	\label{fig:gauss}
\end{figure}

We calculate the 1D line-of-sight velocity dispersion $\sigma_i$ in our simulations as the mass-weighted standard deviation of the velocity along a given axis $i$ in a region between $|z| < 1\,\mathrm{kpc}$. The 3D velocity dispersion is then $\sigma_\mathrm{3D} = \sqrt{\frac{1}{3} (\sigma_\mathrm{x}^2 + \sigma_\mathrm{y}^2 + \sigma_\mathrm{z}^2)}$. We calculate the purely turbulent component of the velocity dispersion that is not affected by thermal broadening. Observationally, a common method for determining the velocity dispersion of a gas is to measure the line-of-sight velocity profile and apply a Gaussian fit to the data. We test how well these two methods agree for our simulations with Fig. \ref{fig:gauss} for an arbitrarily chosen snapshot from the simulation $\Sigma100$.

\section{Turning off stellar feedback}

We run a test model with the highest initial gas surface density while turning off all stellar feedback processes, $\Sigma100$-noFB, for $t - t_\mathrm{SFR} \approx 120\,\mathrm{Myr}$. Except for the stellar feedback processes, the rest of this model is identical to $\Sigma100$ and uses the same initial conditions. With this approach, we want to further quantify the impact of stellar feedback on the velocity dispersion in the WIM. In Appendix Fig. \ref{fig:noFB}, we show the line-of-sight velocity dispersions in the WIM as a function of $\Sigma_\mathrm{SFR}$, similar to Fig. \ref{fig:turbdisp_3D}. Again, we discretise the star formation rate surface density in 0.5-dex-width bins and indicate the mass-weighted thermal sound speed, $c_\mathrm{s}$, in each bin with a grey horizontal line and a 1$\sigma$ standard error as a shaded area. The data are averaged in 5 Myr bins. We want to remind the reader that we drive artificial initial turbulence with a root mean square velocity of $v_\mathrm{rms} = 30\,\mathrm{km\,s^{-1}}$ for the $\Sigma_\mathrm{gas} = 100\,\mathrm{M_\odot\,pc^{-2}}$ models up until star formation sets on and the first sink particle forms.

As discussed above, we find a strong correlation of star formation activity with the velocity dispersions in the WIM, which become supersonic for $\Sigma_\mathrm{SFR} \gtrsim 1.6\times10^{-2}\,\mathrm{M_\odot\,yr^{-1}\,kpc^{-2}}$ (see Section \ref{sec:kinematics}). The Spearman rank correlation coefficients $R_i$, for the $\sigma_i - \Sigma_\mathrm{SFR}$ relation, are $R = (0.89, 0.92, 0.87)$ for the line-of-sight velocity dispersions $\sigma_{x,y,z}$, respectively (see Appendix Table \ref{tab:noFB}). Without stellar feedback, we do not create any kind of outflows and the lack of the two main regulation mechanisms of star formation at high surface densities, i.e. depletion of the star-forming gas reservoir and (primarily early) stellar feedback in the form of hydrogen-ionising radiation, as well as SNe, the star formation rate surface density in $\Sigma100$-noFB is boosted by 0.5 dex compared to $\Sigma100$. In the no-feedback model, there is only a very weak correlation, if any at all, between the velocity dispersion and $\Sigma_\mathrm{SFR}$ with $R_\mathrm{noFB} = (0.25, 0.19, 0.27)$ for $\sigma_{x,y,z}$, respectively, with fairly high $p-\mathrm{values} = (0.20, 0.34, 0.17)$ (compare with $p-\mathrm{values} < 10^{-9}$ for the $\sigma_i - \Sigma_\mathrm{SFR}$ relation including stellar feedback). The maximum vertical velocity dispersion in $\Sigma100$-noFB is about $\sim 15\,\mathrm{km\,s^{-1}}$ higher than in $\Sigma100$. At the same time, the average vertical velocity dispersion in the model that includes stellar feedback for $\Sigma_\mathrm{SFR} > 2.5 \times 10^{-1}$ (the last two $\Sigma_\mathrm{SFR}$ bins in Fig. Appendix \ref{fig:noFB}) is larger with a smaller spread with $\sigma_\mathrm{z,\,high \Sigma_\mathrm{SFR}} \approx (38 \pm 7)\,\mathrm{km\,s^{-1}}$, compared to $\sigma_\mathrm{z,\,high \Sigma_\mathrm{SFR}}^\mathrm{noFB} \approx (28 \pm 16)\,\mathrm{km\,s^{-1}}$. The takeaway is that without stellar feedback and purely by gravitational collapse alone, it is possible to drive turbulence with a high velocity dispersion, however, this does not have to be the case. There are periods in which the gas even drops into the subsonic regime while star formation activity is high. Gravitational collapse together with stellar feedback, on the other hand, generates turbulence which consequently scales with the star formation rate surface density and enters the supersonic regime with velocity dispersions up to $50\,\mathrm{km\,s^{-1}}$ for star formation rate surface densities just below $\Sigma_\mathrm{SFR} \sim 1\,\mathrm{M_\odot\,yr^{-1}\,kpc^{-2}}$. Star formation and the resulting stellar feedback seem to be important constituents of the source of ISM turbulence. It is not possible to disentangle the contribution of stellar feedback alone from the effects of gravity since turning off self-gravity in our simulations, which is principally possible, would result in no star formation at all.

\begin{figure}
	\centering
	\includegraphics[width=.99\linewidth]{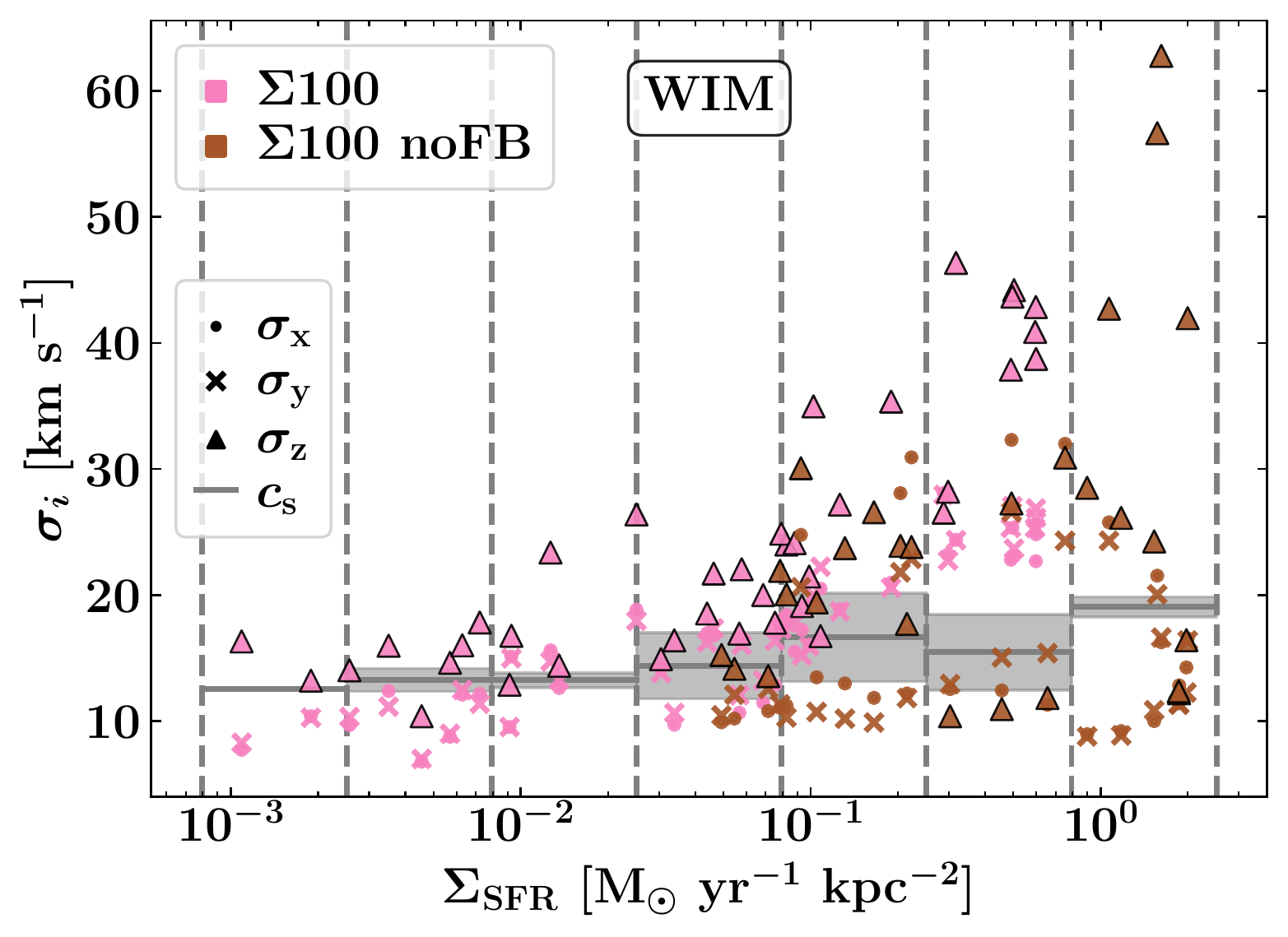}
	\caption{Line-of-sight velocity dispersions in the WIM as a function of $\Sigma_\mathrm{SFR}$ for the highest surface density run $\Sigma100$ and the test-model without stellar feedback $\Sigma100$-noFB. We discretise the star formation rate in 0.5 dex wide bins and indicate the average sound speed $c_\mathrm{s}$ in each of those bins with a grey line horizontal line including a 1$\sigma$ standard error as the shaded area (similar to Fig. \ref{fig:turbdisp_3D}). The data is averaged in 5 Myr bins.}
	\label{fig:noFB}
\end{figure}

\begin{table}
	\centering
	\caption{Spearman rank correlation coefficients, $R$, for the line-of-sight velocity dispersions in the WIM, $\sigma_i$, and $\Sigma_\mathrm{SFR}$ for the highest surface density run $\Sigma100$ and the test-model without stellar feedback $\Sigma100$-noFB.}
	\begin{tabular}{lrrr}
		\hline
		                 & $R_{\sigma_\mathrm{x}}$ & $R_{\sigma_\mathrm{y}}$ & $R_{\sigma_\mathrm{z}}$ \\
		\hline
		$\Sigma100$      & 0.89                    & 0.92                    & 0.87                    \\
		$\Sigma100$ noFB & 0.25                    & 0.19                    & 0.27                    \\
		\hline
	\end{tabular}
	\label{tab:noFB}
\end{table}

\section{Outflow phase structure}
	
Similarly to Fig. \ref{fig:vouts_100}, we present additional plots to quantify the characteristics of the galactic outflow for $\Sigma010$, which we have left out of the main text to not hinder the flow of reading in Appendix Fig. \ref{fig:vouts_010}. We refer the reader to the main text for a detailed discussion.

\begin{figure}
	\centering
	\includegraphics[width=.94\linewidth]{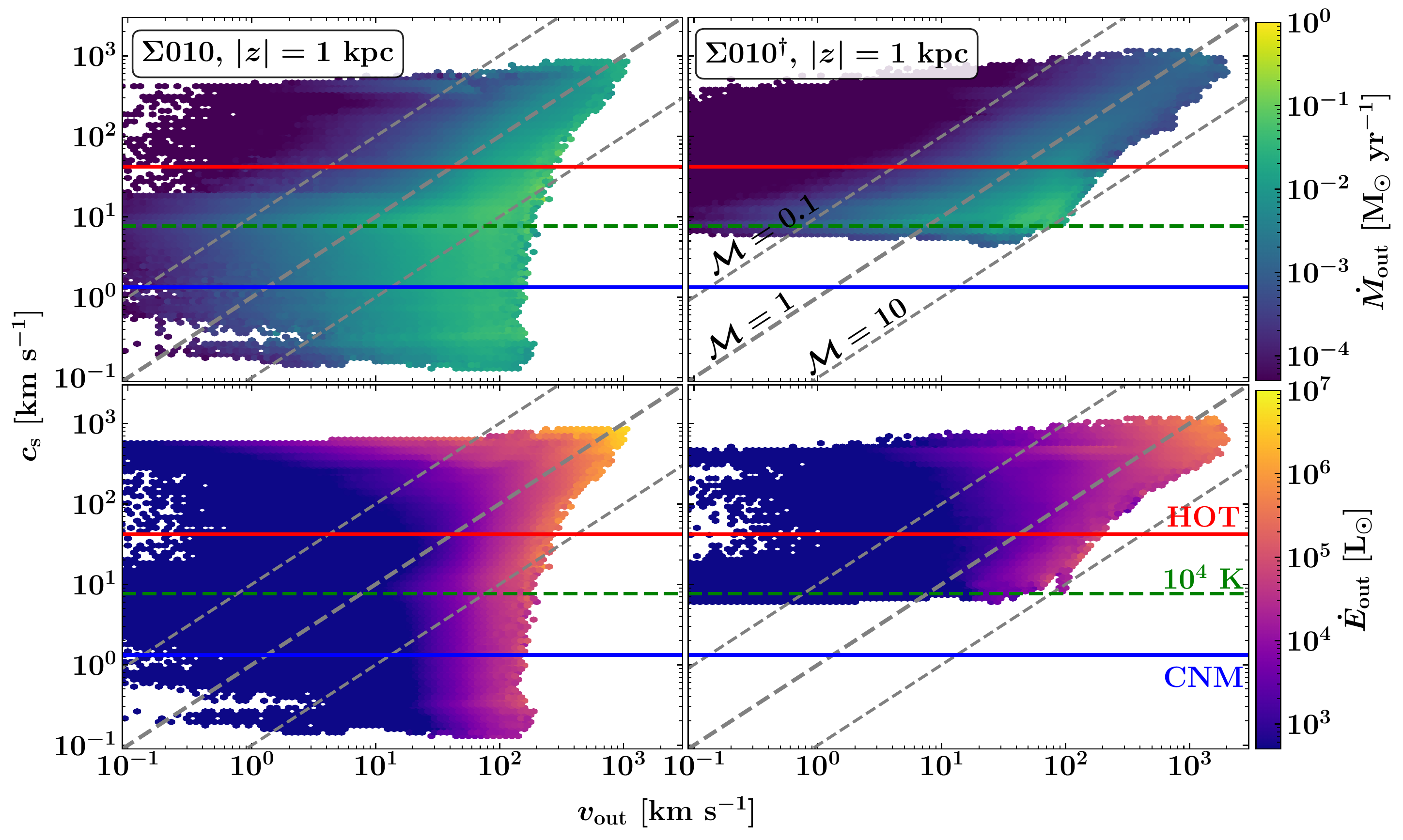}\\
	\includegraphics[width=.94\linewidth]{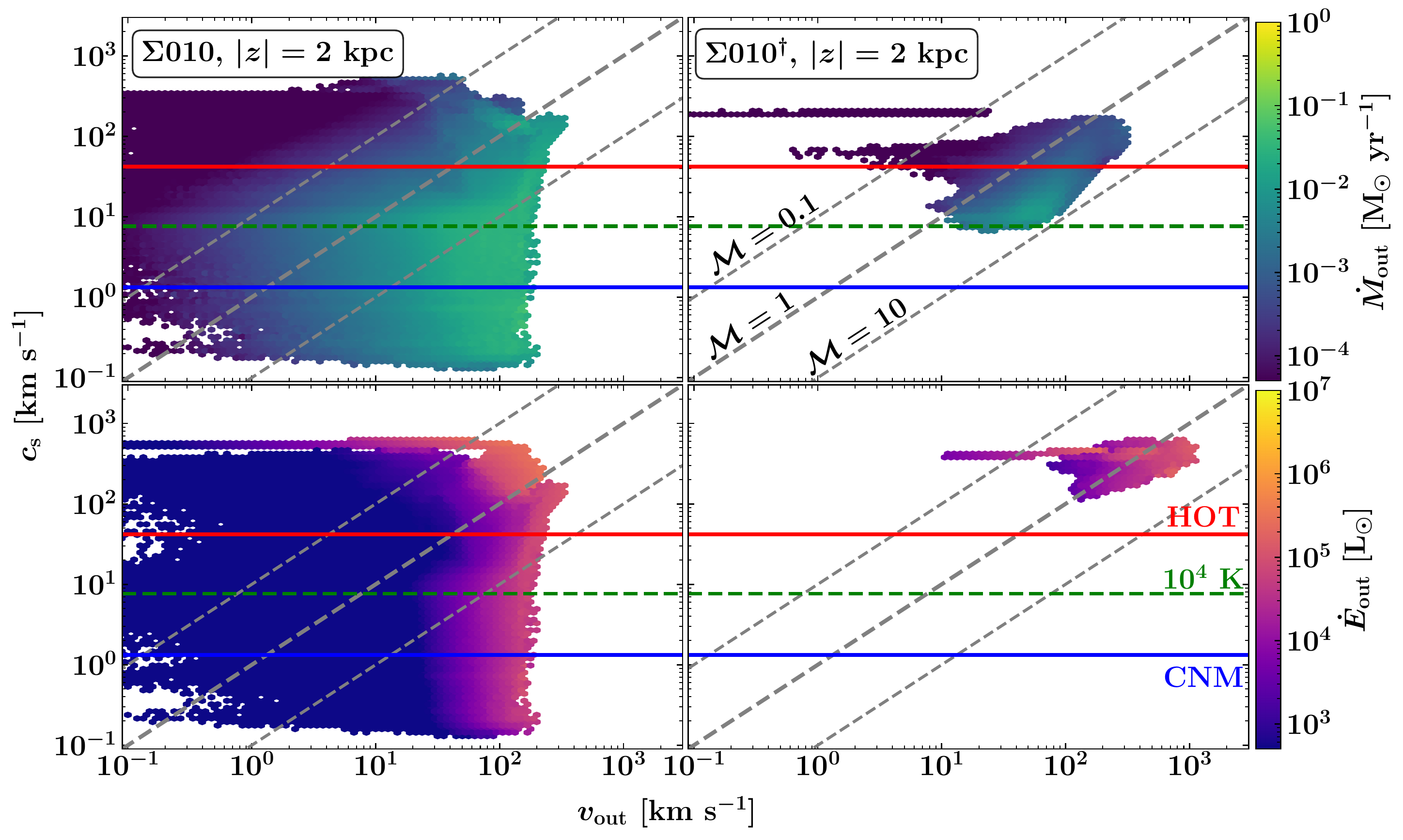}
	\caption{Same as Fig. \ref{fig:vouts_100} but for $\Sigma010$ and $\Sigma010^\dagger$. The outflow is highly multiphase and cannot be described by a single gas phase or characteristic outflow velocity. At a height of $|z| = 1\,\mathrm{kpc}$ the mass outflow is dominated by the warm and cold medium, whereas most of the energy is carried away from the midplane ISM in the hot gas phase. Even though the integrated total mass and energy outflow is comparable for the high surface density runs with and without CRs, the phase structure of the outflow differs drastically with no cold gas component and stronger supersonic outflow velocities in the case without CRs (see also Table \ref{tab:of_phase} and Table \ref{tab:of_velocity}).}
	\label{fig:vouts_010}
\end{figure}
	
\section{Quantitative results}
	
We tabulate the quantitative results of
the analysis of the gas phase structure of the outflow in the Appendix Table \ref{tab:of_phase},
as the sonic structure of the outflow velocities in the Appendix Table \ref{tab:of_velocity} (see Section \ref{sec:PhaseStructureOfTheOutflow});
the average velocity dispersions along the major axis for each $\Sigma_\mathrm{SFR}$ bin of Fig. \ref{fig:turbdisp_3D} in Appendix Table \ref{tab:sigma_sfr}, as well as the gas phase structure of the ISM in Appendix Table \ref{tab:ISM} (see Section \ref{sec:ISMstructure}).

\begin{table*}
	\centering
	\caption{Phase structure of the outflow. Most of the mass is transported out in the warm gas phase and most of the energy is transported out in the hot gas phase. A cold component of the outflow occurs only in the presence of CRs.}
	\begin{tabular}{lccccccccccccccc}
		\hline
		Name                      &      & $\eta_\mathrm{1kpc}$ &      &                        &      & $\eta_\mathrm{2kpc}$ &      &   &      & $\gamma_\mathrm{1kpc}$ &       &   &      & $\gamma_\mathrm{2kpc}$ &      \\
		                          & HOT  & WARM                 & CNM  &                        & HOT  & WARM                 & CNM  &   & HOT  & WARM                   & CNM   &   & HOT  & WARM                   & CNM  \\
		                          & [\%] & [\%]                 & [\%] &                        & [\%] & [\%]                 & [\%] &   & [\%] & [\%]                   & [\% ] &   & [\%] & [\%]                   & [\%] \\
		\hline
		$\mathbf{\Sigma010}$      & 21   & 52                   & 27   &                        & 13   & 54                   & 33   &   & 88   & 8                      & 3     &   & 67   & 21                     & 12   \\
		$\mathbf{\Sigma030}$      & 22   & 61                   & 17   &                        & 26   & 52                   & 22   &   & 88   & 10                     & 1     &   & 91   & 7                      & 2    \\
		$\mathbf{\Sigma050}$      & 23   & 58                   & 18   &                        & 18   & 56                   & 26   &   & 86   & 11                     & 2     &   & 81   & 15                     & 4    \\
		$\mathbf{\Sigma100}$      & 21   & 60                   & 19   &                        & 10   & 62                   & 28   &   & 76   & 20                     & 4     &   & 67   & 23                     & 10   \\
		$\mathbf{\Sigma010^\dag}$ & 25   & 75                   & 0    &                        & 25   & 75                   & 0    &   & 96   & 4                      & 0     &   & 100  & 0                      & 0    \\
		$\mathbf{\Sigma100^\dag}$ & 2    & 89                   & 10   &                        & 3    & 97                   & 0    &   & 49   & 50                     & 1     &   & 38   & 62                     & 0    
		\\
		\hline
	\end{tabular}
	\label{tab:of_phase}
\end{table*}

\begin{table*}
	\centering
	\caption{Sonic structure of the mass-weighted outflow velocity, $v_\mathrm{out}$ and characteristic absolute values of $v_\mathrm{out}$ at the different boundaries. We categorise a velocity with Mach number $\mathcal{M} \leq 1$ as subsonic, $1 < \mathcal{M} \leq 10$ as supersonic, and $\mathcal{M} > 10$ as hypersonic. The characteristic $v_\mathrm{out}$ is calculated as the median $v_\mathrm{out}$ with \nth{25} and \nth{75} percentiles as upper and lower bounds. The percentage of hypersonic outflow and the magnitude of the absolute outflow velocity increase between $|z| = 1\,\mathrm{kpc}$ and $|z| = 1\,\mathrm{kpc}$ for CR-supported outflows. However, only the hottest outflows reach escape velocities high enough to leave the gravitational attraction of a Milky Way-like system (with the Milky Way escape velocity $v_\mathrm{esc} \approx 550\,\mathrm{km\,s^{-1}}$ \citep{Kafle2014}).}
	\begin{tabular}{lcccccccccc}
		\hline
		Name                      &          & $v_\mathrm{out}$ at $|z| = 1$ kpc &            &   &          & $v_\mathrm{out}$ at $|z| = 2$ kpc &            &   &                                &                                \\
		                          & subsonic & supersonic                        & hypersonic &   & subsonic & supersonic                        & hypersonic &   & $v_\mathrm{out}^\mathrm{1kpc}$ & $v_\mathrm{out}^\mathrm{2kpc}$ \\
		                          & [\%]     & [\%]                              & [\%]       &   & [\%]     & [\%]                              & [\%]       &   & [km s$^{-1}$]                  & [km s$^{-1}$]                  \\
		\hline
		$\mathbf{\Sigma010}$      & 7        & 64                                & 29         &   & 6        & 58                                & 36         &   & $27_{17}^{41}$                 & $34_{23}^{47}$                 \\
		$\mathbf{\Sigma030}$      & 6        & 76                                & 18         &   & 18       & 58                                & 24         &   & $34_{21}^{57}$                 & $46_{27}^{73}$                 \\
		$\mathbf{\Sigma050}$      & 6        & 74                                & 20         &   & 8        & 62                                & 30         &   & $47_{28}^{72}$                 & $66_{41}^{102}$                \\
		$\mathbf{\Sigma100}$      & 5        & 74                                & 21         &   & 2        & 65                                & 33         &   & $64_{42}^{95}$                 & $91_{58}^{133}$                \\
		$\mathbf{\Sigma010^\dag}$ & 18       & 82                                & 0          &   & 5        & 95                                & 0          &   & $39_{24}^{56}$                 & $47_{39}^{56}$                 \\
		$\mathbf{\Sigma100^\dag}$ & 1        & 82                                & 17         &   & 1        & 99                                & 0          &   & $66_{40}^{106}$                & $97_{58}^{153}$                \\
		\hline
	\end{tabular}
	\label{tab:of_velocity}
\end{table*}
 
\begin{table}
	\centering
	\caption{Averaged mass-weighted line-of-sight velocity dispersion $\sigma$ in $x-$, $y-$, and $z-$direction with 1$\sigma$ standard error. The $\Sigma_\mathrm{SFR}$ bins (B1-B7) have a width of 0.5 dex and reach from $2.5\times10^{-4}\,\mathrm{to}\,7.9\times10^{-1}\,\mathrm{M_\odot\,yr^{-1}\,kpc^{-2}}$ (see also Fig. \ref{fig:turbdisp_3D}). The velocity dispersion along the outflow axis, $\sigma_\mathrm{z}$, is systematically larger than the other two and increases with $\Sigma_\mathrm{SFR}$ which informs of star formation driven outflows as the source for the high velocity dispersion.}
	\begin{tabular}{lccc}
		\hline
		   & $\sigma_\mathrm{x}$ & $\sigma_\mathrm{y}$ & $\sigma_\mathrm{z}$ \\
		   & [km s$^{-1}$]       & [km s$^{-1}$]       & [km s$^{-1}$]       \\
		\hline
		B1 & $10 \pm 1$          & $9 \pm 2$           & $13 \pm 3$          \\
		B2 & $10 \pm 2$          & $10 \pm 2$          & $13 \pm 4$          \\
		B3 & $12 \pm 4$          & $12 \pm 3$          & $16 \pm 6$          \\
		B4 & $14 \pm 3$          & $15 \pm 3$          & $19 \pm 4$          \\
		B5 & $15 \pm 3$          & $15 \pm 3$          & $20 \pm 6$          \\
		B6 & $20 \pm 3$          & $20 \pm 2$          & $29 \pm 6$          \\
		B7 & $24 \pm 1$          & $25 \pm 1$          & $40 \pm 3$          \\
	\end{tabular}
	\label{tab:sigma_sfr}
\end{table}

\begin{table}
	\centering
	\caption{Characteristic phase structure of the midplane ISM ($|z| = 50\,\mathrm{pc}$). We give the volume filling fraction (VFF) of the warm neutral medium ($300 < T \leq 3\times10^5\,\mathrm{K}$, ionisation parameter $\chi < 0.5$, WNM), the warm ionised medium ($300 < T \leq 3\times10^5\,\mathrm{K}$, ionisation parameter $\chi \geq 0.5$, WIM), and the hot medium ($T > 3\times10^5\,\mathrm{K}$, HOT) and the mass fraction (MF) of the cold neutral medium ($T \leq 300\,\mathrm{K}$, CNM), WNM and WIM.
		The given values are again the global median with \nth{25} and \nth{75} percentiles as upper and lower bounds. There are no systematic trends detectable between the different surface density initial conditions.}
	\begin{tabular}{lccccccc}
		\hline
		Name                      &                & VFF            &                &   &                & MF             &                \\
		                          & WNM            & WIM            & HOT            &   & CNM            & WNM            & WIM            \\
		                          & [\%]           & [\%]           & [\%]           &   & [\%]           & [\%]           & [\%]           \\
		\hline
		$\mathbf{\Sigma010}$      & 43$_{27}^{66}$ & 14$_{9}^{20}$  & 38$_{11}^{53}$ &   & 18$_{14}^{30}$ & 70$_{54}^{77}$ & 8$_{5}^{16}$   \\
		$\mathbf{\Sigma030}$      & 55$_{44}^{75}$ & 12$_{10}^{16}$ & 29$_{5}^{40}$  &   & 11$_{9}^{20}$  & 74$_{66}^{81}$ & 10$_{7}^{15}$  \\
		$\mathbf{\Sigma050}$      & 40$_{23}^{61}$ & 14$_{9}^{21}$  & 33$_{12}^{60}$ &   & 14$_{7}^{19}$  & 66$_{21}^{78}$ & 17$_{9}^{24}$  \\
		$\mathbf{\Sigma100}$      & 41$_{19}^{72}$ & 12$_{7}^{17}$  & 36$_{12}^{64}$ &   & 15$_{9}^{26}$  & 58$_{47}^{81}$ & 18$_{6}^{28}$  \\
		$\mathbf{\Sigma010^\dag}$ & 58$_{38}^{68}$ & 12$_{9}^{14}$  & 30$_{18}^{47}$ &   & 28$_{22}^{36}$ & 66$_{57}^{72}$ & 6$_{4}^{8}$    \\
		$\mathbf{\Sigma100^\dag}$ & 14$_{5}^{64}$  & 7$_{4}^{14}$   & 71$_{27}^{90}$ &   & 24$_{11}^{29}$ & 55$_{47}^{74}$ & 15$_{10}^{26}$ \\
		\hline
	\end{tabular}
	\label{tab:ISM}
\end{table}		
	
\bsp
\label{lastpage}
\end{document}